\documentclass[journal]{IEEEtran}

\usepackage{graphicx}
\usepackage{epstopdf}
\usepackage[cmex10]{amsmath}
\usepackage{amsfonts}
\usepackage[ruled,linesnumbered]{algorithm2e}
\usepackage{algorithmic}
\usepackage{array}
\usepackage{eqparbox}
\usepackage{epsfig}
\usepackage{color}
\usepackage{graphicx}
\usepackage{times}
\usepackage{float}
\usepackage{stfloats}
\usepackage{graphicx}
\usepackage{subfigure}
\usepackage{type1cm}
\usepackage{url}
\usepackage{cite}
\usepackage{enumerate}
\usepackage{lineno}
\usepackage{amsthm}
\usepackage{amssymb}
\usepackage{bbm}
\usepackage{bm}
\usepackage{color}
\usepackage{multirow}
\usepackage{mathtools}
\UseRawInputEncoding
\usepackage{tikz}
\usetikzlibrary{mindmap}  

\usepackage[flushleft]{threeparttable}

\usepackage[short]{optidef}
\allowdisplaybreaks 
\usepackage{enumerate}

\SetKwInput{KwInput}{Input}
\SetKwInput{KwOutput}{Output}

\theoremstyle{remark}

\usepackage{enumitem}
\newlist{tabenum}{enumerate}{1}
\setlist[tabenum]{wide=0pt, 
                  nosep, 
                  leftmargin= * ,
                  label*=\arabic*.,
                  after=\vspace{-\baselineskip},
                  before=\vspace{0.2\baselineskip}
                  }

\begin{document}

\title{Scaling Blockchains with Error Correction Codes: A Survey on Coded Blockchains}

\author{Changlin Yang, \IEEEmembership{Member, IEEE}, Kwan-Wu Chin, Jiguang Wang, Xiaodong Wang, \IEEEmembership{Fellow, IEEE}, Ying Liu, Zibin Zheng, \IEEEmembership{Senior Member, IEEE} 
 %
%
\thanks{C. Yang, Y. Liu and Z. Zheng are with the School of Software Engineering, Sun Yat-sen University, Zhuhai, China (e-mails: yangchlin6, liuy2368, zhzibin@mail.sysu.edu.cn).

K.-W. Chin is with the School of Electrical, Computer and Telecommunications Engineering, University of Wollongong, NSW, Australia (e-mail: kwanwu@uow.edu.au). 

J. Wang is with the School of Computer Science, Zhongyuan University of Technology, Henan, China (e-mail: 2020107237@zut.edu.cn.

X. Wang is with the Electrical Engineering Department, Columbia University, New York, NY, USA (e-mail: wangx@ee.columbia.edu)



%


}}
\maketitle

\begin{abstract}
This paper reviews and highlights how coding schemes have been used to solve various problems in blockchain systems.  Specifically, these problems relate to scaling blockchains in terms of their data storage, computation and communication cost, as well as security. To this end, this paper considers the use of coded blocks or shards that allows participants to store only a fraction of the total blockchain, protect against malicious nodes or erasures due to nodes leaving a blockchain system, ensure data availability in order to promote transparency, and scale the security of sharded blockchains.  Further, it helps reduce communication cost when disseminating blocks, which is critical to bootstrapping new nodes and helps speed up consensus of blocks. For each category of solutions, we highlight problems and issues that motivated their designs and use of coding.  Moreover, we provide a qualitative analysis of their storage, communication and computation cost.
\end{abstract}

\begin{IEEEkeywords}
Blockchain scalability, peer-to-peer networks, error correction codes, distributed systems
\end{IEEEkeywords}

\maketitle

\section{Introduction}
\label{sec_intro}
Blockchain is a decentralized system for building trusts among independent peers.  Its distributed nature helps avoid a single point of failure.  
Further, blockchain offers the following advantages: it ensures data is tamper proof, offers trust and transparency, immutability, maintains an open and global infrastructure that allows any peers to participate \cite{underwood2016blockchain}. 
Consequently, it is of interest to the crypto currency industry, where notable examples include Bitcoin \cite{nakamoto2008bitcoin} and Ethereum \cite{ethereum}.  
Other applications include medical treatment~\cite{mettler2016blockchain}, supply chain management~\cite{BCsupply19}, copyright protection~\cite{MaDRBC}, and file storage~\cite{chen2017improved}.    Further, it can be used to secure the interactions between objects in industrial Internet of Things (IoT) networks, e.g., smart factories~\cite{teslya2017blockchain}. Other examples of blockchain applications in different domains can be found in \cite{BCDomains}.  
In these applications, blockchain is mainly used to (i) track and verify medical treatments, (ii) secure health data, (iii) trace items/ingredients/products, and (iv) protect digital contents, e.g., proof of ownership.
%

\begin{figure}[t]
\includegraphics[width=\linewidth]{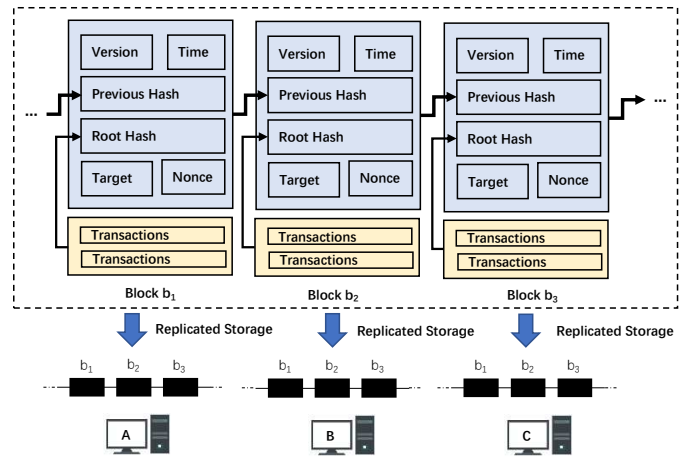}
\centering
\caption{An example of a conventional blockchain, e.g., Bitcoin \cite{nakamoto2008bitcoin}. Each block consists of a block header, colored in blue, and a block body, colored in yellow. The version and time fields in the block header indicate the blockchain version and its generation time. The root hash, e.g., a Merkle tree root, is computed using transactions in a block's body. Each block contains the hash of a previous block, which leads to a chain of blocks.   Further, each block includes a nonce value derived from solving a puzzle such that the hash of the entire block header is less than a target value. This is also known as the Proof-of-Work consensus protocol, where the noun `work' refers to the computation time and/or energy used to solve puzzles.  }
\label{traditional_BC}
\end{figure}

%
A key feature of blockchain is that data is organized into blocks, see Fig.~\ref{traditional_BC} for an example.  These blocks form a hash chain, whereby each block is protected by a hash value that also includes the hash value of the previous data block.  Hence, any change to a block affects the hash value of subsequent blocks. 
Another feature is its use of a consensus protocol, such as  proof-of-work (PoW) \cite{nakamoto2008bitcoin} and practical byzantine fault tolerance (PBFT) \cite{castro1999practical} to ensure it is computationally impractical for an attacker to modify transactions in a block as this would entail re-calculating a hash chain. 
A key concern when applying blockchain is its data storage requirement.
Specifically, all blocks are replicated and stored in fully functional nodes, aka {\em full nodes}.  As a result, full nodes require a large storage space. For example, in Bitcoin \cite{nakamoto2008peer}, the storage requirement of a full node is approximately 380 GB at the beginning of 2022, which is an increase of 20\% as compared to its size in 2021 \cite{bitcoinsize}. Another example blockchain is Ripple~\cite{Ripple2019}, where its storage size grows by 12 GB per day, and its entire blockchain size exceeds 14 TB as of 2022~\cite{ripple_size}.   Consequently, the significant storage cost required by blockchain limits its scalability, and leads to fewer full nodes or centralization~\cite{raman2021ToN}.

Henceforth, much research effort has been dedicated to reducing the storage requirement of blockchain. To date, popular approaches include: 
\begin{itemize}
    \item {\em Light node}~\cite{nakamoto2008bitcoin} -- The main idea is to have two types of nodes.  In particular, so called {\em light nodes} stores only block headers.   
    On the other hand, {\em full nodes} store all blocks in a blockchain.
    To verify a transaction is in a given block, a light node uses Simplified Payment Verification (SPV) \cite{nakamoto2008peer}, where it relies on a full node to verify blocks. 
    \item {\em Pruned blockchain}~\cite{nakamoto2008bitcoin} \cite{nadiya2018block} \cite{kim2019scc} -- Its main idea is to remove blockchain information that is not required for block miners. 
    In Bitcoin, miner checks whether Unspent Transaction Outputs (UTXOs) are spent in previous blocks. Hence, spent UTXOs can be deleted without affecting a new block.  Alternatively, miners only store the balance of each account \cite{nadiya2018block} to ensure a consumer has sufficient coins, or require users to store proofs that they own assets or coins \cite{Mina}. 
    \item {\em Sharding blockchain}~\cite{zamani2018rapidchain} -- Its main idea is to divide a blockchain into multiple sub-chains. Each sub-chain is an independent blockchain that is operated by a group of nodes or {\em community}; this is also called a {\em sharding}. A node only belongs to a single community, meaning it only needs to store a given fraction of a blockchain. Hence, node storage requirement is proportionally reduced with  increasing number of shards.   In fact, it scales logarithmically according to the number of communities.
    %
\end{itemize}
The aforementioned approaches, however, raise new security issues. In particular, a {\em light node} can be deceived by malicious nodes that pose as full nodes~\cite{karame2016bitcoin}.   Further, as the ratio between full and light nodes increases, the resulting blockchain has a semi-centralized structure, which erodes the decentralized property of blockchain~\cite{perard2018erasure}.
{\em Pruned blockchains} discard information permanently, and thereby it cannot be used for applications that require ready access to historical information, e.g., medical records \cite{mettler2016blockchain}. 
A {\em sharding blockchain} requires cross-chain verification and merger of sub-chains, which increase the functional complexity of nodes. Moreover, the number of nodes in each shard decreases with increasing number of shards. This reduces the security of a sub-chain, and thereby degrades the reliability of blockchain \cite{das2018security}.
To this end, {\em coded} blockchain is a promising approach to reduce its storage requirement and also retain its security features.  It applies research from distributed storage \cite{Dimakis2011Survey}, where it uses an error correction code to encode blocks and stores these blocks in a distributed manner. Hence, each node only stores some coded {\em fragments}, which helps reduce its data storage requirement as well as communication cost~\cite{CachinIDV}.   Advantageously, a coded blockchain achieves storage reduction without permanent information loss.  That is, when a node or user needs uncoded data, e.g., for verifying a transaction or some records, it collects coded fragments and runs a decoding algorithm to restore original blocks.  Moreover, each node independently performs encoding/decoding, which retains the distributed nature of blockchain.  In addition to storage reduction, the error correction feature of codes can detect and correct data altered by malicious nodes.    As a result, coded blockchain is poised to address both scalability and security issues of blockchain.   Lastly, as we will see in Section~\ref{sec_light}, coding helps thwart data availability attacks.

\begin{figure}[t]
\includegraphics[width=\linewidth - 3em]{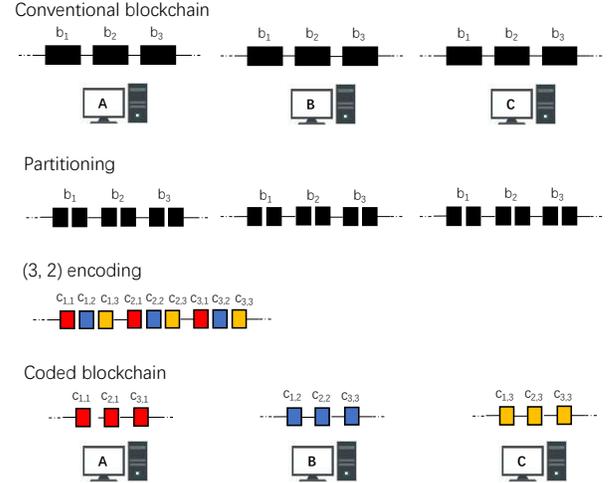}
\centering
\caption{An example of coded blockchain procedures. Each of the original blocks $b_1, b_2, b_3$ (in black) can be partitioned into two fragments, when then encoded into three coded fragments $b_{x, 1}, b_{x, 2}, b_{x, 2}$ where $x = 1, 2, 3$, represented in red, blue and yellow, respectively. In a coded blockchain system, instead of storing entire blocks, nodes store a coded fragment, which halves the storage requirement of nodes. }
\label{fig: 1.1}
\end{figure}

Fig.~\ref{fig: 1.1} shows an example of coded block storage.  There are three blocks in the blockchain. In a conventional blockchain system, each node stores a copy of the entire blockchain. By using a coded blockchain, a node partitions a block into two fragments and encodes them using for example a $(3, 2)$ Reed-Solomon code \cite{1994Reed}. This results in three coded fragments for each block. Then each node only stores a coded fragment of each block, which halves its storage requirement.  Further, this reduces communication cost as it reduces the amount of data that is sent to other nodes requiring a coded fragment.
This example also illustrates the trade-off between data availability storage space, i.e., a node has all the necessary blocks to verify transactions versus the case where it needs to download coded blocks first before verifying transactions.


%
\begin{table}[htbp]
    \centering
    \caption{A comparison of Blockchain storage reduction methods. }
    \begin{tabular}{|m{1.7cm}|>{\centering}m{1.3cm}|>{\centering}m{1.1cm}|>{\centering}m{1.1cm}| >{\centering\arraybackslash}m{1.3cm}|}
    \hline
          & {\bf Distributed} & {\bf Integrity} & {\bf Security} & {\bf Storage}   \\
       \hline
        Light node \cite{nakamoto2008bitcoin}  & X    & X & X & 0.05\% \cite{Ethereum_light_node} \\
        \hline
         Pruned \cite{nakamoto2008bitcoin}  & \checkmark   & X & \checkmark & 2\% \cite{pruned_node_space} \\
        \hline
         Sharding \cite{zamani2018rapidchain} & \checkmark    & \checkmark & X & 6.1\% \cite{zamani2018rapidchain} \\
        \hline
         Coded \cite{dai2018low} & \checkmark    & \checkmark & \checkmark & 0.47\% \cite{wu2020distributed}\\
        \hline
        
    \end{tabular}
    \label{tab_compare}
\end{table}
Table~\ref{tab_compare} compares a coded blockchain with light nodes, pruned and a sharding blockchain. 
Specifically, it compares them according to the following features: (i) {\em distributed}, i.e., they do not rely on a central entity,  (ii) {\em integrity}, meaning their data is temper proof and unabridged, (iii) {\em security}, where nodes participate in consensus to maintain overall safety, and (iv) {\em storage}, whereby their storage requirement is compared with traditional blockchains such as Bitcoin and Ethereum.
From the table, we see that a light node does not participate in consensus nor store block data. Hence, it not distributed, and does not maintain the integrity and security of blockchain. A pruned node is able to independently verify new transactions, but permanently deletes information, meaning it is unable to  maintain integrity. A sharding blockchain dedicates a few nodes to each shard, which raises its security risk.  
In contrast, a coded blockchain does not have these limitations as it retains the strengths of conventional blockchain but requires nodes to have a significantly smaller storage space.
%

%
%
%

%
%
%
To date, there are two major lines of research into coded blockchain.  As shown in Table~\ref{tab_history_temp}, most works focus on distributed storage of blockchain. Their aim is to use error correction codes to reduce the storage requirement of nodes.  In this respect, researchers have studied secure storage, communication cost, boostrapping cost and distributed computing of coded blockchains.  Another research direction aims to defend against Data Availability Attack (DAA), see details in Section \ref{sec_light}.  In this respect, a key issue is that light nodes may readily a block with fraudulent transactions from a malicious node.  
To this end, coded blockchain helps because even though a malicious node may delete or tamper with coded fragments, honest nodes will be able to reconstruct the original data.  Further, it provides a method to ensure transparency whereby all data is always available for verification by any nodes.
\begin{table*}[htbp]
    \centering
    \caption{Development of Coded Blockchain}
    \begin{tabular}{|>{\centering}m{3cm}|>{\centering}m{1.5cm}|>{\centering}m{1.5cm}| >{\centering\arraybackslash}m{1.5cm}|>{\centering}m{1.5cm}|>{\centering}m{1.5cm}| >{\centering\arraybackslash}m{1.5cm}|}
    \hline
          & {\bf Pre. 2017} & {\bf 2018} & {\bf 2019} & {\bf 2020} & {\bf 2021} & {\bf 2022}   \\
      \hline
        Distributed Storage  & -   & \cite{dai2018low} \cite{perard2018erasure} & \cite{quan2019transparent}&   \cite{wu2020distributed}& \cite{li2021lightweight} & \cite{qin2022downsampling} \\
        \hline
         Secure Distributed Storage & \cite{raman2017dynamic}    & \cite{raman2018distributed} \cite{raman2018dynamic}  & \cite{kim2019efficient}  &   \cite{mesnager2020threshold} \cite{qi2020bft}  & \cite{raman2021ToN} & \cite{9771862} \\
        \hline
         Distributed Storage and Communication   &  \cite{HBadger}  & \cite{cebe2018network} \cite{BEATBFT}& \cite{choi2019scalable}\cite{chawla2019velocity} \cite{8922597} &  \cite{2020GCBlock} & \cite{9609913} \cite{yang2021storage} & \cite{DLedgerTse} \cite{zhang2022speeding} \\
        \hline
        Distributed Storage and Bootstrapping  &  -  & - & \cite{kadhe2019sef} \cite{mitra2019patterned}& \cite{pal2020fountain} \cite{gadiraju2020secure} & \cite{tiwari2021secure} & - \\
        \hline
        Distributed Storage and Computing  &  -  & \cite{polyshard2018} & - & \cite{2020PolyShard} & \cite{wang2021low} \cite{khooshemehr2021discrepancy}\cite{sasidharan2021private} \cite{asheralieva2021throughput} & \cite{wang2022JSIT} \\
        \hline
        Data Availability Attack  &  -  & \cite{al2018fraud} & - &  \cite{yu2020coded}\cite{cao2020cover} \cite{sheng2020aced} & \cite{mitra2021StopSet} \cite{mitra2021overcoming} \cite{mitra2021communication} & \cite{santini2022optimization} \cite{mitra2022polar} \cite{battaglioni2022data} \\
        \hline

    \end{tabular}
    \label{tab_history_temp}
\end{table*}

\begin{figure}[t]
\includegraphics[width=0.9\linewidth]{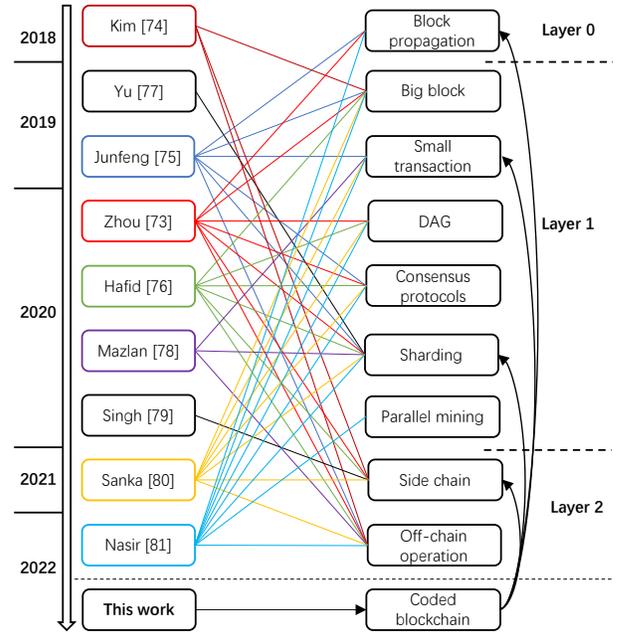}
\centering
\caption{Comparison with existing surveys (left) and their reviewed solutions (right). We use blockchain layers defined in \cite{zhou2020solutions}. }
\label{compare_survey}
\end{figure}

\subsection{Contributions}
To the best of our knowledge, this is the {\em first} survey on coded blockchain research.
To date, a number of surveys have focused on non-coding based approaches for scaling blockchains, see in Fig. \ref{compare_survey}.   For example, the works such as~\cite{zhou2020solutions, kim2018survey, 8823874, hafid2020scaling} review scaling solutions relating to off-chain operation, side chain, sharding, consensus protocols and directed acyclic graph (DAG). On the other hand, the work in~\cite{yu2020survey} focuses on sharding.  
In \cite{mazlan2020scalability}, the authors discussed solutions that scale healthcare blockchains. 
In \cite{SideChain2020}, the authors review works on sidechains.
The work reported in~\cite{sanka2021systematic} \cite{nasir2022scalable} systematically analyzes research for aforementioned blockchain scalability solutions, with additional discussion on parallel mining. 
However, these works do not discuss {\em coded blockchain}.
In particular, coded blockchain can be implemented to improve the performance of existing blockchain propagation, storage, sharding and side chain solutions. 
Hence, this survey fills this critical gap.

\begin{figure}[t]
\includegraphics[width=\linewidth]{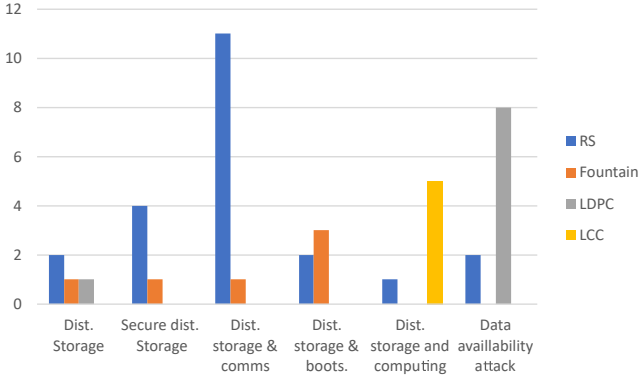}
\centering
\caption{The number of works that implement RS, fountain, LDPC and LCC codes to address scalability issues in blockchains. }
\label{count_code}
\end{figure}

Referring to Table \ref{tab_history_temp},  this paper introduces the use of error correction codes to reduce the blockchain storage, communication cost and bootstrapping cost. It also discusses the ability of such codes on distributed computing and against data availability attacks in blockchain.  
Note that in this survey, we do not consider coded network storage, e.g., \cite{5550492}, which assumes a synchronous environment, because in blockchain systems, nodes operate asynchronously and maybe malicious.
Common error correction codes include Reed-Solomon (RS) \cite{1994Reed}, Low Density Parity Check (LDPC) \cite{2002Low}, Luby Transform (LT) \cite{2002LT}, Raptor \cite{2006Raptor}, Lagrange Polynomial \cite{2020PolyShard} codes to name a few. 
Fig.~\ref{count_code} shows the number of works using such codes to address scalability issues in blockchain. We see that most works implement Reed-Solomon code, and LDPC codes are popular as well due to their efficient decoding process.
%
%


        

\begin{figure}[t]
\includegraphics[width=\linewidth]{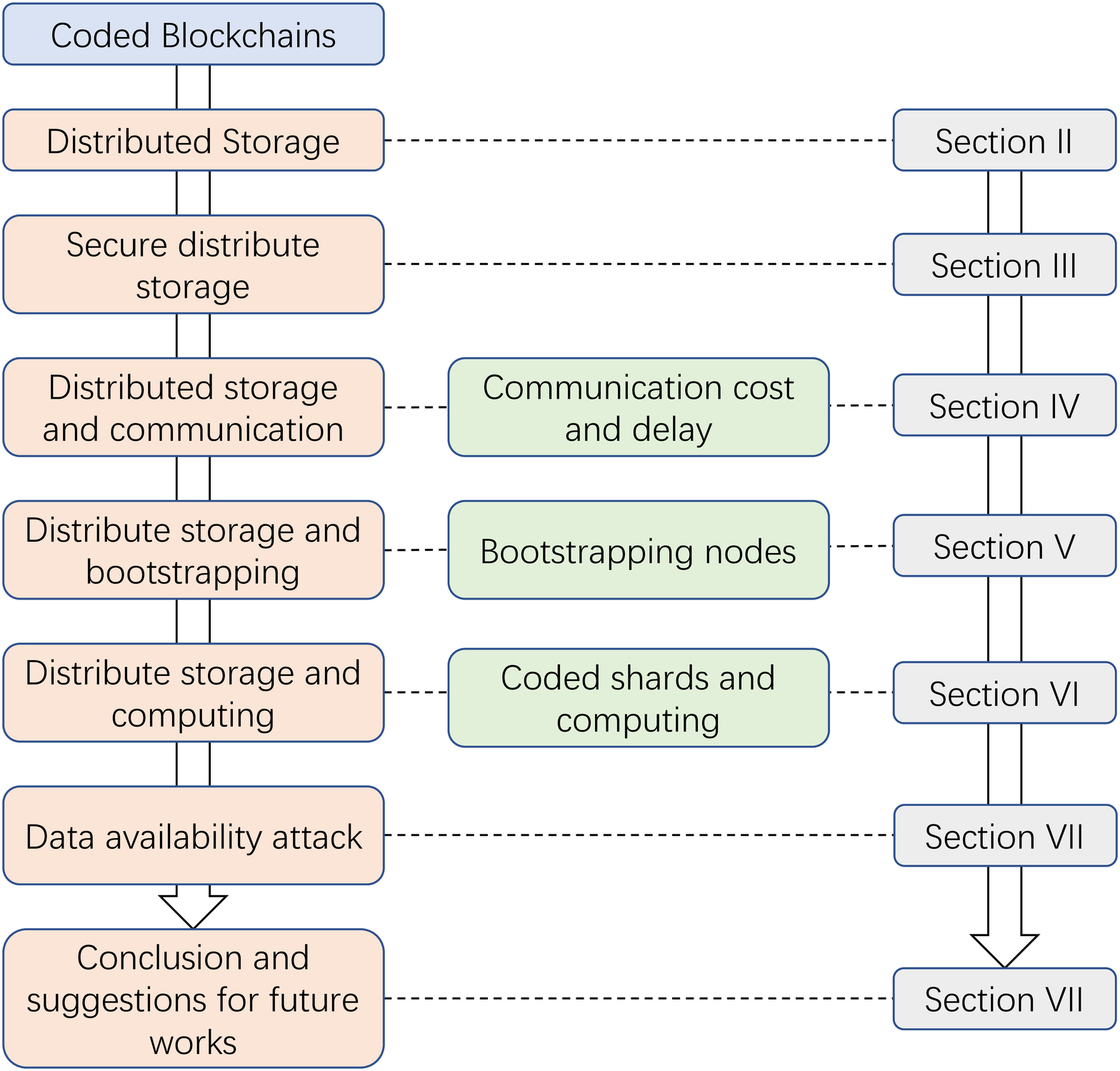}
\centering
\caption{The structure of this survey. }
\label{fig_subsections}
\end{figure}
The structure of this survey is shown in Fig \ref{fig_subsections}.  Next, we review works that employ various coding schemes in an effort to reduce storage cost.  Then in Section III, we consider works that ensure data integrity, confidentiality and prevent denial of service attacks.  Section IV reviews works that aim to efficiently propagate and retrieve coded blocks.  Section V presents solutions that aim to bootstrap nodes quickly.   Then in Section VI, we review works that aim to scale sharded blockchains.  Section VII highlights works that aim to ensure data availability by using various coding schemes to discourage malicious nodes from hiding transactions.
Lastly, Section VIII provides a qualitative comparison of all prior works and presents some future works.
%

\section{Distributed Storage}
The approaches that aim to employ coding to reduce network storage borrow ideas from distributed network storage~\cite{Dimakis2011Survey} and build on the seminal work by Rabin~\cite{RabinIDA}, which is then extended by Cachin et al.~\cite{CachinIDV} to consider byzantine faults.
Briefly, coding helps reduce the total size of a blockchain as follows.  Assume a blockchain has storage size $\Omega$, and we are given parameter $k$ and $n$, meaning we have a code rate of $\frac{k}{n}$.  For example, we can encode $k$ fragments using Reed-Solomon codes \cite{1994Reed} to create $n$ coded fragments.  These coded fragments are then distributed to the $\Gamma$ participants of a blockchain.  Ideally, each participant only stores a coded fragment of size $\frac{\Omega}{k}$; i.e., the storage requirement is now $\frac{1}{k}$ of the total amount $\Omega$. 
For example in Fig.~\ref{fig: 1.1}, we have $k=2$ and $n=3$ such that each participant only stores $\frac{1}{2}$ of entire blockchain.
To reconstruct the blockchain, a participant requires $k$ blocks, which can be retrieved from other participants.
As noted in Dai et al. \cite{dai2018low}, the $k$ fragments can also be coded using a rateless code such as LT~\cite{2002LT}.
Another approach to reduce storage is via downsampling and coding of transactions. In \cite{qin2022downsampling}, the authors consider storing only those blocks with the highest entropy.  Specifically, the {\em entropy} of blocks relates to the amount of information, e.g., balances of participants, they carry with regards to the current blockchain state.  To further reduce storage requirement and to ensure recovery of original blocks, they consider encoding transactions.  In particular, each node stores a coded fragment that is calculated using a number of transactions or UTXOs following the robust Soliton distribution of LT codes~\cite{2002LT}.
Full nodes usually store all blocks or an entire blockchain.   However, Wu et al.~\cite{wu2020distributed} propose to have full nodes store one of a given number of blocks.  The authors also propose to encode a group of blocks at a time using erasure codes.  In addition to reducing storage requirement, this also helps a new node to quickly acquire blocks from a small number of participants, and thus reduces communication cost.  Further, they propose to use LDPC, which helps reduce computational complexity.  In contrast, the work in \cite{perard2018erasure}  uses random linear codes, which requires matrix inversion.  Moreover, the use of LDPC helps optimize Galois field size and number of codewords used to encode a group of blocks.
To date, prior works aim to address a number of issues.  
First, as highlighted in \cite{dai2018low}, the value of $n$ must be chosen with respect to the number of participants.  This becomes an issue if a blockchain has time varying number of participants.   This issue is addressed by adopting a rateless code.  The downside, however, is that it may require a large field size, which hinders its use by nodes with limited energy and computational resources.
Similarly, in \cite{wu2020distributed}, the number of symbols used to encode a block and the number of valid codewords have an impact on the number of erased symbols that impede the recovery of a block.  Further, these parameters have an impact on network load, i.e., the number of participants to be contacted in order to recover a block. 
Apart from that, as noted in Perard et al.~\cite{perard2018erasure}, the value of $k$ and the number of coded fragments maintained by a node has an impact on communication overheads and amount of storage required by a node.  It is worth noting that the method in \cite{wu2020distributed} incurs a smaller storage size than the approach proposed by Perard et al.  Further, it impacts the availability of a block, i.e., whether there are $k$ coded fragments that can be recovered to re-construct a block.  
In this respect, Li et al. \cite{li2021lightweight} consider a $(n_i, k_i)$ Reed-Solomon codes, where block $i$ is divided into $k_i$ segments, which are then used to produce $n_i$ coded fragments.  These coded fragments are then stored at $n_i$ participants. They propose to jointly optimize the value of $n_i$, placement of $n_i$ coded fragments, and also the load or number of requests sent to participants.
The second issue is the distribution of coded fragments among participants.  For example, Dai et al. \cite{dai2018low} assume all participants store coded fragments.  On the other hand, Perard et al.~\cite{perard2018erasure} introduce low-storage nodes that store some coded fragments and hash of a block; each low-storage node creates these coded fragments independently.  Their motivation is to encourage nodes with limited resources to participate in blockchain.  This in turns allows a blockchain system to have more participants and also to alleviate the bandwidth requirement of full nodes, which store all blocks.
%
Advantageously, the number of coded fragments stored by low-storage nodes is determined by their storage capacity and the age of a block.  Specifically, a low-storage node may remove some coded fragments of a block.
%

%
In general, the said prior works mainly consider (i) the availability of coded fragments for reconstruction/decoding, (ii) computational complexity.  Both of which are affected by the parameters of a code, e.g., size of a finite field or/and number of symbols used to encode one or more blocks.  
Referring to Table~\ref{tab_para}, the storage and recovery probability reduce with increasing value of $k$. Moreover, the recovery probability of a block increases with the finite field size $q$ for finite field $\mathbb{F}_q$. In addition, the processing complexity increases with $k$, $n$ and the size $q$ of finite field $\mathbb{F}_q$. The communication overhead increases with $k$ because a node needs to obtain more coded fragments to decode. 
Further, the type of coding schemes, e.g., random linear codes versus LDPC, will have an impact on computational complexity. Hence, the choice of coding schemes will have an impact on resource constrained devices.
Apart from \cite{li2021lightweight, wu2020distributed}, not many works emphasize communication cost or network load.  This becomes a critical issue if devices have a bandwidth and energy constraints, meaning they may not have the resources to retrieve all coded fragments to decode a block.  Further, the chosen coding scheme and its associated parameters may result in network congestion; i.e., the network load may scale with the number of low-storage nodes.  We will discuss this issue further in Section~\ref{BOOT}.
Lastly, in addition to the above issues, these prior works ensure their proposed method does not reduce the integrity and decentralized nature of blockchain.  Further, they ensure their method is robust against attacks.
%
%

%

\begin{table}[htbp]
    \centering
    \caption{Effect of coding parameters. }
    \begin{tabular}{|m{0.8cm}|>{\centering}m{1.1cm}|>{\centering}m{1.2cm}|>{\centering}m{1.2cm}| >{\centering\arraybackslash}m{1.3cm}|}
    \hline
          {\bf Param.} & {\bf Storage} & {\bf Recover probability} & {\bf Processing complexity} & {\bf Comms. overheads}   \\
       \hline
        $k$  & $\mathcal{O}(\frac{1}{k})$    & $\mathcal{O}(\frac{1}{k})$ & $\mathcal{O}(k)$ & $\mathcal{O}(k)$ \\
        \hline
         $\mathbb{F}_q$  & -  & $\mathcal{O}(q)$ & $\mathcal{O}(q)$  & - \\
        \hline
         $n$ & -    & - & $\mathcal{O}(n)$ & - \\
        \hline
        
    \end{tabular}
    \label{tab_para}
\end{table}

\section{Secure Distributed Storage}
In a blockchain system, key security issues of concern include data integrity, confidentiality and denial of service (DoS) attacks.
Blockchain ensures {\em data integrity} via the use of hash chains.  Specifically, each block is protected by a hash value and also that of the previous block.  Hence, any changes to one block will require an attacker to re-compute the hash of subsequent blocks in the chain~\cite{nakamoto2008bitcoin}.
{\em Confidentiality} relates to encryption of blocks, where privacy is of concern to block chain applications such as healthcare~\cite{9445631}.  In this respect, key management is an issue and whether a compromised node results in an attacker obtaining all data or the secret key used to encrypt blocks.  Further, active adversary is a concern, whereby a malicious node seeks to modify the content of one or more blocks.
Lastly, an {\em DoS attack} aims to stop a participant from recovering blocks or transactions, or to delay the propagation of blocks or transactions to nodes.
Next, we discuss works that consider coded storage solutions that address the above issues.
Readers interested in non-coding based solutions are referred to \cite{Lock1, Lock2Survey, 9093015}.  Note that, some works such as \cite{8957108} have used blockchain to ensure the security of an existing distributed storage system.  They are out of the scope of this survey.
A number of works have considered secure distributed storage of blockchains.  An interesting set of works are by Raman et al.~\cite{raman2017dynamic,raman2018distributed,raman2018dynamic,  raman2021ToN}, where they consider an archival ledger.
Further, they consider {\em cold} and active transactions in an archival ledger.  They define {\em cold} transactions as completed transactions no longer in use or accessed infrequently.  For example, these transactions may correspond to buying and selling of properties, where a transaction contains proof of ownership.
They offer a number of solutions to the aforementioned security issues.
Their blockchain system uses zones, where the participants of each zone have permission to access a block via a private key.  Further, they store coded fragments of a block encrypted using the said private key.  A key innovation is that the private key is coded using Shamir's key sharing scheme~\cite{Shamir1979how}; briefly, this scheme constructs a code for the private key, where coded fragments are stored at different participants.  To recover the private key, a node must retrieve $k$ of the said coded fragments.
In this respect, Raman et al. showed that if a certain number of participants in a zone are corrupted, then a transaction may be compromised.  To this end, they propose to construct a new zone membership over time such that each participant is a peer of every other participants eventually; i.e., the zones form a {\em chain} over time.  This constructs ensures that there are sufficient number of honest participants in each zone.  It also requires an adversary to corrupt all nodes, not just those in a zone, before it can compromise a transaction.  As a result, data integrity improves with more participants.
A key drawback of the scheme proposed by Raman et al.~\cite{raman2021ToN} is that transactions or blocks are replicated across zones.  To this end, the authors of~\cite{9771862} apply fountain codes to first encode a number of blocks before applying a secret sharing scheme, e.g., \cite{9174266}, to each coded block.  After that, the encrypted and coded blocks are distributed to $m$ zones.  To decode and decrypt a block, a user request $d$ coded encrypted blocks from each zone.  In this respect, a key concern is secret sharing schemes that are efficient in terms of communication bandwidth usage during repair and decoding, e.g., \cite{8006842}.
Advantageously, the use of fountain codes ensures a constant finite field size is used by a secret sharing scheme, which helps with decoding.
Another drawback of the work in~\cite{raman2021ToN} is its high communication cost when a node fails.  This is because a node failure leads to a repair process that requires communication with multiple peers in another zone.  To this end, Kim et al.~\cite{kim2019efficient} propose a hierarchical secret sharing scheme comprising of a global and local secrets.   Shares of the global secret are distributed across zones, whilst each zone has a local secret shared by some number of nodes in that zone.
Advantageously, their scheme allows a single node failure to be recovered locally, i.e., from other nodes that belong to the same zone as the failed node.

Mesnager et al. \cite{mesnager2020threshold} propose a secure threshold verifiable multi-secret sharing scheme based on Feldman's verifiable secret sharing \cite{feldman1987practical}. Their scheme is inspired by \cite{raman2021ToN}. In particular, they first encrypt each block using the AES-256 symmetric key encryption algorithm. They then encode the encrypted block using a Reed-Solomon code and store them in a distributed manner. They also state that their scheme is post quantum secure since there is no traditional and quantum attacks against it. They show that the proposed scheme has a lower recovery communication cost than  \cite{raman2021ToN} and \cite{kim2019efficient}. 
%

Lastly, the work in \cite{qi2020bft} considers the requirement of PBFT when distributing fragments coded using Reed-Solomon codes.  Specifically, they aim to ensure a sufficient number of coded fragments are stored on a given number of honest nodes.
Qi et al. \cite{qi2020bft} encode the blockchain with coding parameters such that it can be recovered when $1/3$ nodes are malicious. However, the number of malicious nodes may increase when new nodes join the network. To ensure their system robust against Byzantine attack, they propose a replication scheme, where each coded chunk is replicated onto some number of nodes.  Further, they propose a re-encoding mechanism. In particular, when a node joins the network, a leader node is elected to recover blockchain and re-calculate coding parameters. The leader node then sends the recovered blockchain and revised parameters to all nodes, where each of them re-encodes blocks.
%

%

\section{Communication Cost and Delays \label{sec_comms} }
Blockchain is a peer-to-peer network that relies on a gossip protocol~\cite{9539193} to propagate transactions and blocks to peers/neighbors~\cite{nakamoto2008bitcoin}.  
Briefly, a node first advertises its blocks and transactions, along with their hash, to its neighbors.  These neighbors then request and download missing blocks or transactions from the node.   They then verify downloaded blocks and transactions before advertising their availability.
%
%
%

\begin{figure}[t]
\includegraphics[width=0.8\linewidth]{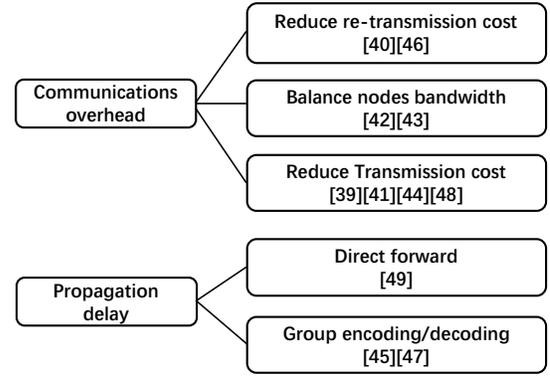}
\centering
\caption{Summary of works for communication cost and propagation delays in Section \ref{sec_comms}. These works fall into five categories and aim to reduce communication overhead or propagation delays in blockchains. }
\label{comms_summary}
\end{figure}

There are two key concerns, see Fig.~\ref{comms_summary}. The first concern is {\em communication overhead}. 
The propagation of transactions/blocks in a blockchain may result in duplicates and congestion.  These result in inefficient bandwidth usage, where the underlying physical network routes/transmits the same block or transactions many times.
Further, the use of a consensus algorithm such as PBFT~\cite{miguel1999pbft} incurs a high signaling overhead.  For example, the message complexity of PFBT is $\mathcal{O}(N^2)$, where $N$ is the number of participants.  
Hence, a key research aim is to reduce the amount of communication bandwidth when propagating blocks to participants.

The second concern is {\em propagation delay}, which is a function of two factors: transmission time of a block, and its verification time. Note that transmission time involves time to  advertise, request and to download a block.   Propagation delay causes blockchain forks \cite{6688704}, meaning there are inconsistencies in the blockchain that affect the security of a blockchain system~\cite{Gobel2017Increased}.  In this respect, proof-of-work computation must be longer than the maximum propagation delay to ensure blocks have a chance to propagate to all nodes~\cite{wang2022JSIT}.    Apart from that, large propagation delays may result in orphan blocks, meaning a node has wasted its computational resources on a block that is rejected by other nodes~\cite{8922597}.
To this end, the key aim of prior works is to reduce the factors that contribute to the propagation delay of a block.

\begin{figure}[t]
\centering
\subfigure[Conventional blockchain rely on replication.  When a block transmission fails, the transmitter needs to re-transmit an entire block to the receiver.]{%
\includegraphics[width=0.7\linewidth]{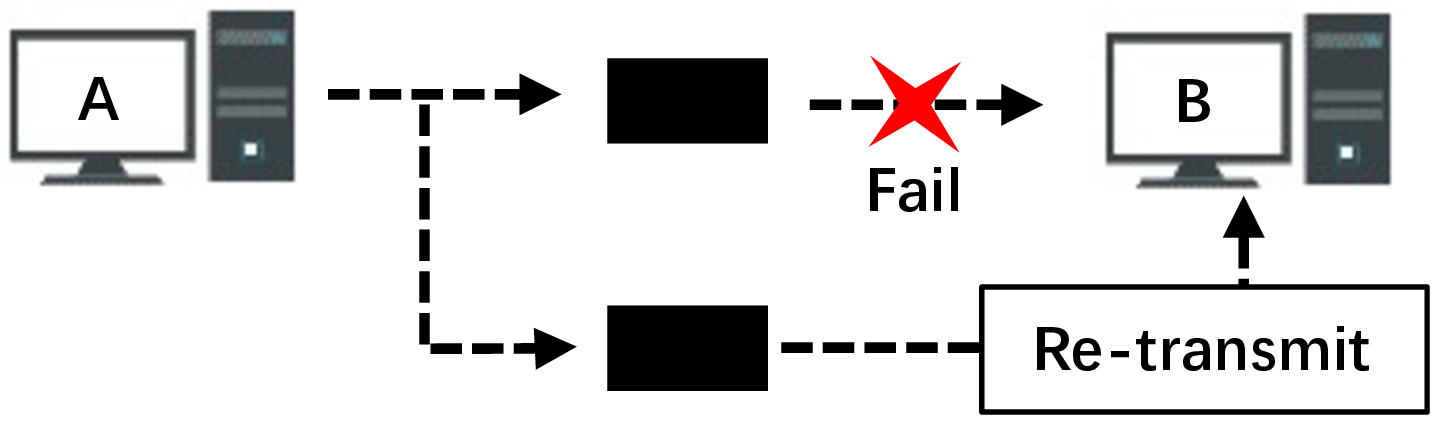}
\label{comms1}}
\subfigure[Coded block propagation.  When a coded fragment transmission fails, the transmitter only needs to re-transmit a coded fragment rather than an entire block.]{%
\includegraphics[width=0.8\linewidth]{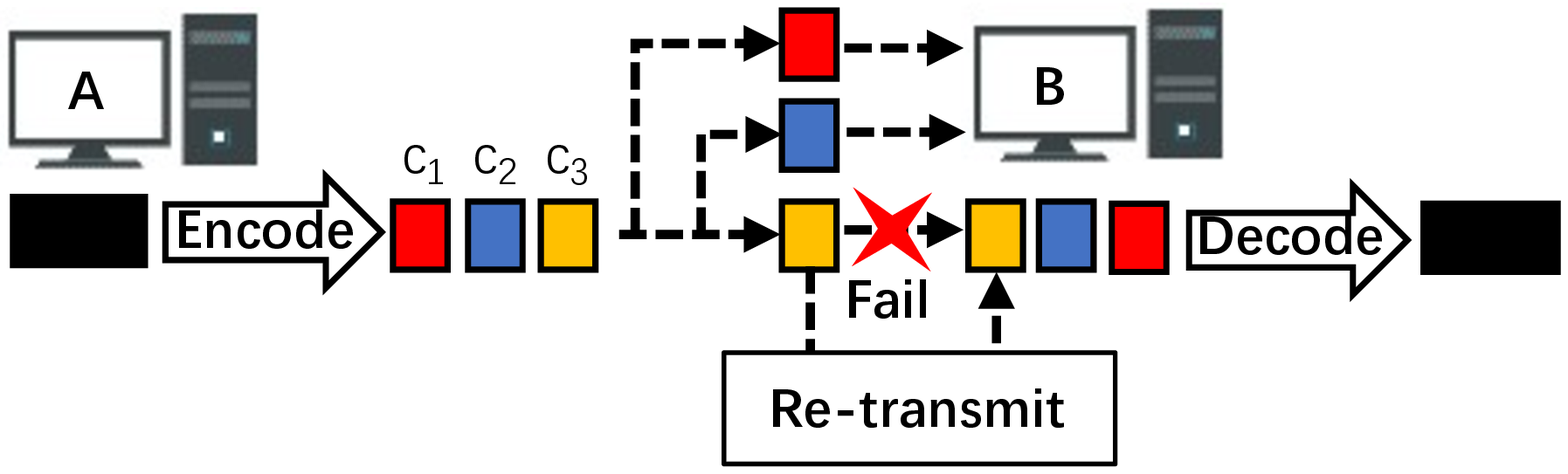}
\label{comms2}}
\subfigure[Network coded block propagation.  The transmitter sends randomly combines and send coded fragments to relays $R_1$, $R_2$ and $R_3$. Then the receiver is able to decode the block from any of these relays. The receiver also acts as a relay by sending a random combination of received coded fragments. ]{%
\includegraphics[width=0.8\linewidth]{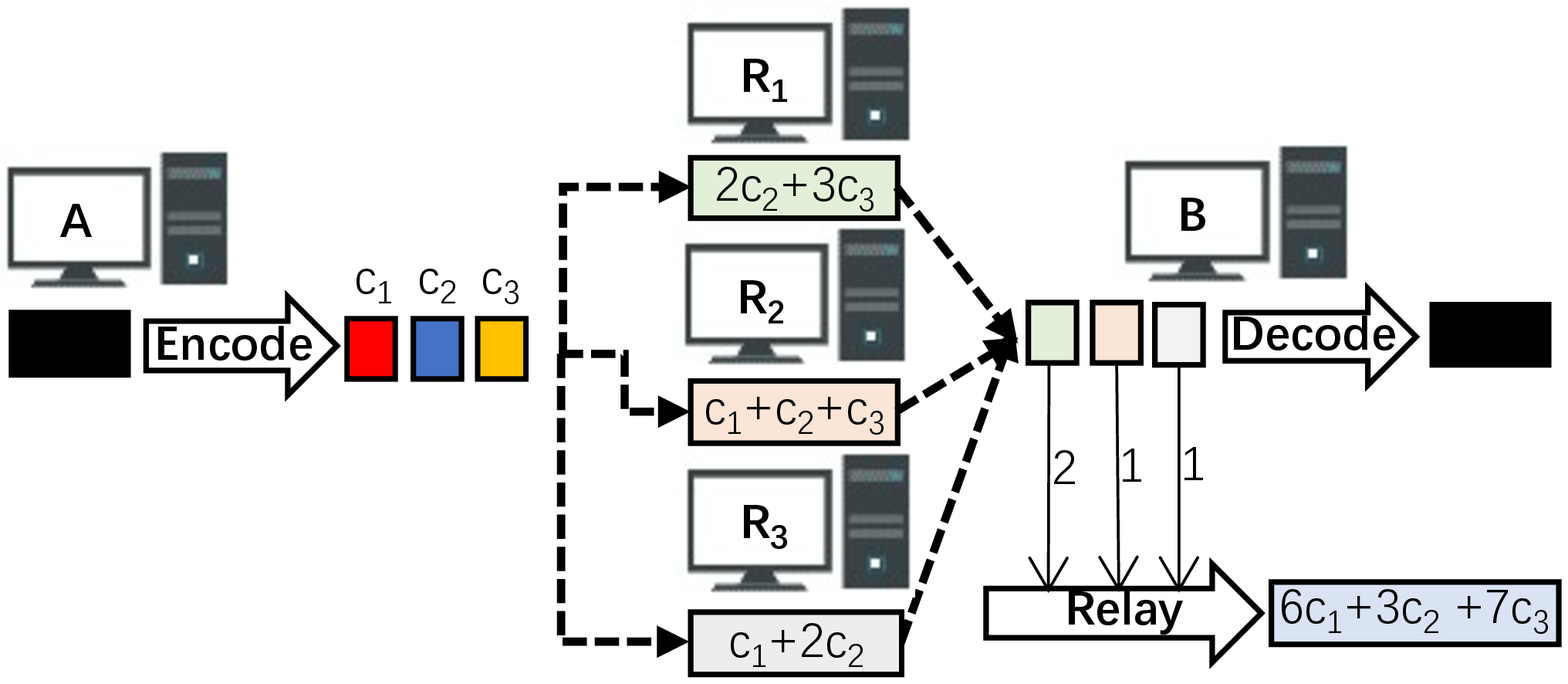}
\label{comms3}}
\quad
\caption{Block propagation of conventional, coded and network coded blockchains. }
\label{comms}
\end{figure}

\subsection{Block Propagation}
A promising approach to improve the dissemination of blocks is via coding.
Fig.~\ref{comms} shows examples of replication, coded and network coded block propagation. We see that coded block propagation, see Fig.~\ref{comms2}, reduces re-transmission cost when there is a transmission failure as compared to replication based block propagation, see Fig.~\ref{comms1}. 
Fig.~\ref{comms3} shows an example of network coding.  In particular, a sender sends random linear combinations of fragments.  A receiver is then able to decode and reconstruct the original block from sufficient number of such fragments.
From the Fig.~\ref{comms3}, we see that with network coding, the link from each relay to the receiver only needs to transmit $\frac{1}{3}$ of the entire block, which reduces the maximum bandwidth requirement.
This advantage thus motivate the use of network coding to improve different protocols and applications. For example, in \cite{5061931}, it is used to improve the throughput of a transport protocol, and in \cite{5462030}, it is used for video streaming.

To date, two works have considered exploiting network coding to reduce the communication cost of PBFT; their approach is similar to the transport and video streaming examples above.
In~\cite{9609913}, Braun et al. study how random linear network coding improves the different phases of PBFT.  They consider nodes (replicas) that communicate over multi-hop paths or intermediate nodes.   Consequently, these intermediate nodes are able to apply network coding. Their simulation results show that network coding helps improve the prepare and commit phases of PBFT.  Moreover, increasing number of replicas and having larger block sizes further improve performance.  Lastly, the number of intermediate nodes has an impact on time taken for receivers to receive all blocks. 
%

Another work is carried out by Cebe et al. \cite{cebe2018network}.  The authors consider a permissioned blockchain system operated by resource constrained wireless devices.  These devices are connected in a mesh; i.e., they have a direct connection to one another.  A transmitter encode blocks into {\em generations}, where each generation consists of a number of fragments.   A transmitter then encodes each fragment and broadcast coded fragments to other devices.  Receivers then decode these coded fragments.  Once they have managed to decode all fragments in a generation, they send an acknowledgment message to the transmitter.  The transmitter stops transmission once all generations have been acknowledged by receivers.  In addition, they consider retransmission policy, e.g., timeout, of a generation.  This is important because devices operate over wireless links.
%
%

Unlike the above works, Chawla et al.~\cite{chawla2019velocity} apply rateless erasure codes.
%
Specifically, a block is encoded using a fountain code by a node.  It is then advertised to neighbors of the said node.  Each neighbor then proceeds to request symbols of the block from their neighbors.  A node decodes the block once they have received sufficient number of coded symbols.
Advantageously, nodes do not become bottlenecks as symbols are downloaded from different neighbors.

%

%
Recently, Jin et al.~\cite{8922597} consider erasure coding and clustering. Their goal is to ensure all participants have the same set of transactions. Their method creates clusters of participants. Each cluster has a leader.  A sender who wishes to send transaction(s) to participants send the ID of these transactions to each leader.  The leader then forwards transaction IDs to cluster members, which in turn reply with the ID of missing transactions.  This information is relayed back to the sender.  It then creates Reed-Solomon coded blocks and send them to leaders.  The participants of each cluster then recover missing transactions using these coded blocks.

\begin{figure}[t]
\centering
\subfigure[In conventional Byzantine consensus, node $A$ transmits a block to all nodes in the pre-prepare stage. Then each node validates the block.  Once verified, a node broadcasts the block to other nodes in the prepare stage. Lastly, if a node receives more than $f+1$ prepare messages, it broadcasts a commit message to other nodes.  Consensus is reached if at least $2f+1$ nodes are honest. This incurs a communication overhead of $\mathcal{O}(Bn^2)$, where $B$ is the block size.]{%
\includegraphics[width=0.9\linewidth]{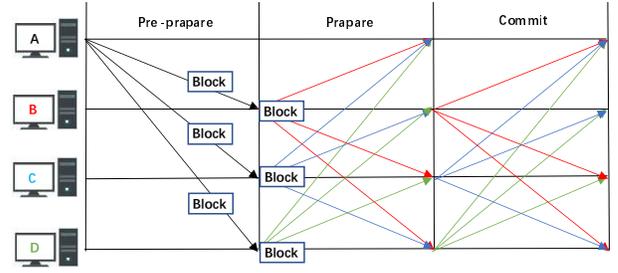}
\label{PBFT_origin}}
\subfigure[By leveraging error correction codes, the block producer $A$ encodes the block into coded fragment using $(n-2f, n)$ Reed-Solomon codes, e.g., two original fragments and two parity fragments, and sends each fragment to one other node. It also sends a Merkle tree proof to show the correctness of the coded fragments~\cite{HBadger}.  After receiving a coded fragment, a node validates it and broadcasts it if is correct. Lastly, upon receive four fragments, each node recovers the original block.  It is able to correct up to $f$ error fragments \cite{1994Reed}. This incurs a communication overhead of $\mathcal{O}(\frac{B}{n-2f}n) = \mathcal{O}(Bn)$. ]{%
\includegraphics[width=0.9\linewidth]{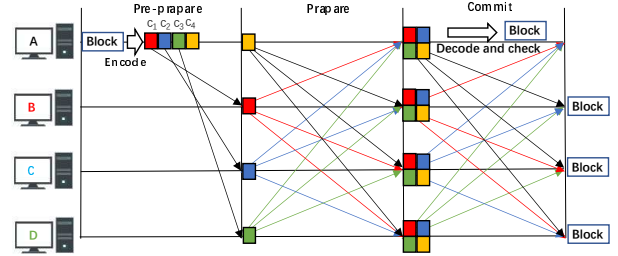}
\label{PBFT_code}}

\caption{Block propagation in conventional and coded Byzantine consensus. There are $n=4$ nodes and the number of tolerable Byzantine nodes is $f=1$. }
\label{PBFT_prop}
\end{figure}

Lastly, we note that a number of works have used erasure codes to reduce the communication or broadcast cost of consensus protocols in the presence of byzantine faults.
A notable effort is HoneyBadgerBFT~\cite{HBadger}, which aims to improve throughput.   To this end, it equips the reliable broadcast protocol of \cite{Bracha84} with an erasure code in order to reduce its communication cost or the size of the protocol's {\em initial} and {\em echo} message. Further, nodes only transmit the hash of coded fragments.  This process can be seen in Figure~\ref{PBFT_code}.
Note that BEAT \cite{BEATBFT} offers variants of HoneyBadgerBFT that are optimized for various scenarios.
Recently, reference~\cite{DLedgerTse} notes that the time to reach consensus can be delayed if there are straggler nodes; i.e., nodes with a low bandwidth.  A key concern is that these nodes require a significant amount of time to download and verify a block.  To this end, they apply the verifiable information dispersal protocol~\cite{CachinIDV} to reduce the amount of data downloaded by each node.  Specifically, a node with a new block applies erasure coding on the block and sends coded fragments/segments plus their respective hash to other nodes.  As each coded fragment is small, straggler nodes are able to download these coded fragments quickly.  Further, this allows a node to check for data availability as it can download pieces from other nodes to reconstruct a block.
Choi et al.~\cite{choi2019scalable} also consider the bandwidth of nodes.  Specifically, they propose a general scheme to distribute coded blocks to nodes.   It generalizes sharded blockchain~\cite{luu2016secure} when partitioning a blockchain to further consider minimizing the maximum bandwidth between nodes when running PBFT~\cite{miguel1999pbft} amongst a group of nodes.  They apply network coding~\cite{ahlswede2000network} where they design a coded block assignment scheme that ensures nodes have some number of coded fragments that correspond to its available bandwidth.  Advantageously, their scheme does not require nodes to carry out decoding first in order to verify a block.
\subsection{Propagation Delay}
To date, there are two key approaches that aim to minimize the propagation delay of a block.  The {\em first} approach aims to speed up the time taken by a node to verify and propagate a block.  The {\em second} approach considers placing required coded blocks within some number of hops away from participants.

Zhang et al. \cite{zhang2022speeding} consider the first approach.  Their approach also uses the compact block relay protocol~\cite{9219639}, which helps reduce communication bandwidth via payload reduction or compression.  They note that the method in~\cite{chawla2019velocity} uses a store-and-forward protocol, meaning a node is required to decode a block before propagating it to the next node.  To this end, Zhang et al. propose a cut-though coded block propagation scheme that takes advantage of rateless codes, which allows a node to download coded fragments from different neighbors.    A key advantage of their approach is that when a receiver node receives a coded fragment, it directly propagates this coded fragment to the next node. Hence, the decoding and propagating are carried out simultaneously. 
Another approach to reduce communication cost is to store coded fragments within some number of hops away from a node.
To this end, Yang et al. \cite{yang2021storage} proposed a $Q$-hop localization problem, which calls for a solution to ensure coded fragments are available for download from a node located Q-hops away. 
%
Critically, these coded blocks must allow for the reconstruction of any blocks.  To this end, they propose an integer linear program (ILP) to determine the assignment of coded fragments to nodes.
Further, they outlined an encoded block transmission routing algorithm to download a needed coded fragment from the nearest node.  They also proposed a distributed algorithm that allows nodes to independently decide whether to store coded blocks.  
Similarly, Qu et al. \cite{2020GCBlock} propose to group nodes that are close to each other in terms of their physical distance.  Each group has a {\em duty} or leader node that is responsible for creating coded fragments.  Further, the leader is responsible for detecting and banning malicious nodes; i.e., nodes that refuse to respond to request for coded fragments.  They use a fractional repetition code~\cite{Rouayheb2010Fractional} to deal with dynamic group membership.  In particular, the code allows its parameters to be adjusted easily as nodes leave and join a group.


\section{Bootstrapping Nodes}\label{BOOT}
When a new node joins a blockchain system, it needs to obtain the latest blockchain state. This process is called node bootstrapping.   This means a new node is required to download the entire blockchain from one or more full nodes. 
However, as noted in \cite{kadhe2019sef}, the storage size/cost of full nodes is considerable, which places a significant strain on full nodes in terms of communication and computation cost to bootstrap new nodes.   
In particular, full nodes become congestion points, and thus they limit the creation of new full nodes, which in turn has an impact on the decentralization property of blockchain.
%

%
To this end, the following research questions are of interest: (i) how does a newly joined node obtain all blocks or the current blockchain state without overwhelming existing full nodes?  (ii) how does a solution protect against nodes that inject malicious blocks?  (iii) how to reduce communication cost? In this respect, a key performance metric is bootstrap cost, which is defined as the number of {\em honest}, as opposed to adversarial, full nodes to be contacted in order to recover all blocks.

Next, we discuss how approaches based on fountain codes and repair codes address the aforementioned questions.
%
\subsection{Fountain Codes}
The main idea is for full nodes to apply fountain codes on validated blocks.  This also means they only need to store coded blocks, which reduces their storage requirement.  A new full node then collects $k$ coded fragments from some subset of full nodes.  This thus helps reduce the amount of traffic directed at a full node.
Note the decoding process can be carried out using the iterative peeling decoding process~\cite{2002LT}. 

The work in~\cite{kadhe2019sef} uses the above idea to reduce the storage of full nodes.  Their results show that a full node with 191.48 GB of data can be reduced to only 195.6 MB. Further, they consider adversarial nodes that inject tampered coded blocks.  To address this issue, the authors propose to incorporate the hash of decoded blocks in the iterative decoding process. This helps avoid propagating a tampered coded block onto subsequent iterations when decoding future blocks.  
A key limitation, however, is that the method in \cite{kadhe2019sef} assumes a new node has access to correct block headers in order to verify the correctness of decoded blocks. However, these headers may have been corrupted by an attacker. 
To solve this problem, Pal \cite{pal2020fountain} proposes to add a verification field in each block. This verification field contains a certificate signed by a random selected committee to guarantee the integrity of a block. A new node checks both hash and verification field of a block during decoding. If one of the hash and verification fields are not consistent, it discards the coded block. 

%
Another limitation with the method in~\cite{kadhe2019sef} is that its decoding complexity increases logarithmically with the number of input blocks~\cite{tiwari2021secure}.   
To this end, the authors of \cite{tiwari2021secure} employ Raptor codes~\cite{2006Raptor}, which have almost constant decoding complexity per block.  Briefly, the proposed encoding process has two layers. In the first layer, a full node encodes raw blocks using an LDPC code to obtain some intermediate coded blocks.  The second layer uses an LT code.  Advantageously, due to the said intermediate coded blocks, the LT code at the second layer can use a less complex degree distribution.   However, as shown in \cite{tiwari2021secure}, this results in a higher bootstrap cost as compared to \cite{kadhe2019sef}.
To bootstrap a node, it first obtains some coded fragments from {\em honest} full nodes.  It then applies the said iterative peeling decoder. Once it has sufficient number of decoded intermediate blocks, it applies an LDPC decoder to obtain raw blocks. Advantageously, if some intermediate blocks are not decodable in the LT code layer, they can be treated as erasures and corrected in the LDPC layer. Hence, the method in \cite{tiwari2021secure} has a higher probability of decoding success than the method in~\cite{kadhe2019sef}.
%

\subsection{Repair Codes}
Recall that fountain code based methods require a new node to first download coded fragments and decode the entire blockchain before encoding the raw blocks to create new coded fragments. This incurs a high bootstrapping or communication cost~\cite{gadiraju2020secure}. It is also inefficient because the new node discards all the downloaded coded fragments used to obtain raw blocks.
%

One approach to improve communication efficiency is by applying regenerating codes~\cite{5550492}.  These codes are known for their bandwidth efficiency. 
Briefly, given $k$ fragments, we generate $\Delta k$ fragments and encode them into $n\alpha$ pieces.  Then we distribute these pieces to $n$ devices, where each device stores $\alpha$ coded pieces. We can contact any $k$ devices to recover the $k$ fragments.  Further, as part of its {\em node repair} process, a regenerating code has a parameter $d$.  A new device contacts $d$ other devices and retrieves $\beta < \alpha$ coded pieces from each device.  The new device then stores $\alpha$ of the  $d\beta$ downloaded pieces.
The parameter $\Delta$ and $\alpha$ have a direct impact on the storage requirement of devices and also communication cost.

Applying the idea of regenerating codes, Gadiraju et al. \cite{gadiraju2020secure} improve RapidChain~\cite{zamani2018rapidchain}, a sharded blockchain system.  
Their aim is to reduce the communication cost of new nodes joining shards.  To do this, their scheme considers a new participant joining a shard as being equivalent to node repair in regenerating codes, where it contacts $d$ participants in a shard to determine its $\alpha$ coded pieces.  
The encoding of blocks in a shard is carried out using a Vandermonde matrix~\cite{kalman1984generalized}.  This allows easy addition of nodes, whereby after a new node is added into a shard, the code remains a minimum bandwidth regenerating code.
Apart from that, Gadiraju et al. consider $p$ malicious nodes.  Consequently, a bootstrapping node contacts $d+2p$ devices to ensure it recovers its coded fragments correctly.  Further, a shard re-configures its coding parameters at each time epoch to account for larger block sizes.
%
%

%
An interesting approach is to take advantage of node failure patterns.  Specifically, erasure codes are traditionally optimized for {\em all} possible erasure patterns.  However, Mitra et al. \cite{mitra2019patterned} observe that the uptime of nodes in a blockchain system exhibits periodic patterns  In other words, some coded fragments are lost periodically.  This means an erasure code can be optimized to only account for these erasure patterns. To this end, Mitra et al. designed a a coded fragment repairing algorithm that uses Reed-Solomon codes, where its parameters are optimized to repair erasures caused by known (patterned) node failures. 
\begin{figure}[t]
\centering
\subfigure[In conventional sharding blockchains, each shard maintains an independent sub-chain. They use cross-shard transactions to exchange information among shards. However, a shard is compromised if a majority of nodes in a shard are malicious.  In this example, an attacker is able to manipulate a shard by controlling three nodes.]{%
\includegraphics[width=0.9\linewidth]{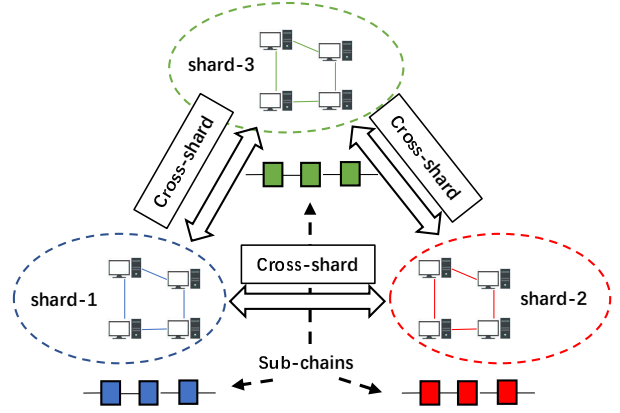}
\label{shard_tradition}}
\subfigure[In a coded sharding blockchain, e.g., \cite{polyshard2018}, when a shard generates a block, it calculates a polynomial over blocks from all shards in a distributed manner. Then each shard stores the chain of polynomials instead of an independent sub-chain. This ensures the same security level as conventional un-sharded blockchains, e.g., Bitcoin \cite{nakamoto2008bitcoin}. ]{%
\includegraphics[width=0.9\linewidth]{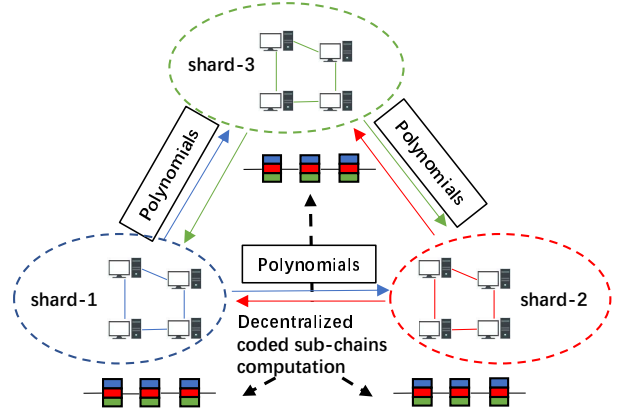}
\label{shard_coded}}

\caption{Comparison between conventional sharding and coded sharding blockchains. }
\label{shard_BC}
\end{figure}

%
%
\section{Coded Shards and Computing}
Sharding or partitioning is a well-known technique used to scale databases~\cite{SQLBook}.   To this end, Luu et al.~\cite{luu2016secure} apply it to permissionless blockchains with the aim to improve their transaction rates.  Critically, they are the first to develop a secure sharding protocol, called ELASTICO, that considers byzantine nodes.  The key idea is to partition nodes into smaller committees, where each committee, aka sub-chain, is responsible for a part of the blockchain; i.e., the transactions managed by committees are disjoint.  
Each committee runs a byzantine consensus algorithm, e.g., PBFT~\cite{miguel1999pbft}, to determine a valid shard.  It then signs and sends the valid shard to a final or consensus committee.  Members of the final committee also run a consensus algorithm to determine (i) that the shard from each committee is signed correctly, and (ii) to agree on a final hash value that is computed over all shards.   Further, it creates a set of random values for use by all committees in their proof-of-work for the next shard.  The final committee then broadcasts the union of all shards, corresponding hash value, and random values to all participants.
%

%
One concern with ELASTICO~\cite{luu2016secure} is that if there a few nodes in each shard, then it is vulnerable to an attack, see Fig.~\ref{shard_tradition} for an example.  Specifically, as noted in \cite{polyshard2018}, in order for ELASTICO to scale with the number of nodes, more sub-chains or committees will have to be created.  However, this reduces the number of committee members, meaning an entire shard could be compromised if a majority of  committee members are dishonest.
To this end, coded sharding or Polyshard~\cite{polyshard2018,2020PolyShard} aims to scale sharded blockchain and retain the security provided by blockchain.
%
Each node stores a coded sub-chain that is calculated using all other sub-chains.  Specifically, the coded sub-chain, which allows data to be corrected even when an attacker controls a whole shard, see Fig.~\ref{shard_coded} for an example.  To do this, Polyshard employs a balance based verification mechanism, whereby a new block is broadcasted to all shards. Each node in each shard then verifies the correctness of the new block through Lagrange coded computing (LCC)~\cite{yu2019lagrange}. Next, all nodes exchange their verification results to achieve consensus.
A key advantage of Polyshard is that it has the same storage requirement as ELASTICO but provides the same security level as conventional un-sharded blockchain.  Further, as noted in \cite{wang2022JSIT}, the use of LCC means that there is no longer any need to re-assign or rotate committee members; it thus solves the risk of having a majority of dishonest committee members.
%

%
%
Inspired by Polyshard, Wang et al. \cite{wang2021low, wang2022JSIT} propose a coded sharding scheme that enables cross shard transactions and low block propagation latencies.  Note that Polyshard does not support inter-shard transactions. Wang et al. propose a two-dimension shard, where transactions inputs and outputs are divided into shards.  For each block, the inputs of all transactions are divided into incoming strip shards, and outputs are divided into outgoing strip shards.  LCC~\cite{yu2019lagrange} is then applied to each input and output strip shard. Since they partition transactions into inputs and outputs, cross shard transactions, where their input and output belong to different shard, can be verified. In addition, Wang et al. propose a coded propagation scheme to reduce propagation latencies due to limited bandwidth. Specifically, each community selects a leader to first exchange outgoing and incoming strips to form a whole block. These leaders then propagate the entire block within its community. 

%
There are two other works that seek to exploit LCC to improve blockchain. 
In~\cite{asheralieva2021throughput}, Asheralieva et al. consider block generation in resource restricted devices, e.g., an Internet of Things (IoT) device with limited computational power and bandwidth. They divide blockchain nodes into miners and validators. Miners generate block using LCC~\cite{yu2019lagrange} and validators check newly generated blocks. When generating a block, miners encode the block into LCC coded segments and send them to validators. After a validator receives sufficient number of coded segments, it decodes and obtains a raw block and verifies it. It then encodes the verification result using LCC and broadcasts it to other validators.  Finally, all validators decode the verification results and accept the block if there is consensus. 
%

%
There are also works that consider issues relating disparity of information and anonymity.
Specifically, the aim is to create differential information between honest and malicious shards, which interrupts honest shards from performing LLC decoding. 
In~\cite{khooshemehr2021discrepancy}, Khooshemehr et al. introduce a {\em differential} attack on PolyShard \cite{polyshard2018}.  Recall that in PolyShard, the new block generated by each shard is broadcasted to the entire network for LCC decoding. An attacker can control a shard, makes it a malicious shard by compromising a small number of nodes. This is achievable since there is only a few nodes in each shard. Then an attacker sends the new block generated by malicious shards only to other malicious shards. Further, when these malicious nodes received a coded fragment from an honest shard, they can send a wrong verification result. 
To preserve anonymity, conventional blockchain uses anonymous accounts.  Another approach is to employ private information retrieval protocols~\cite{PQuery2001}.  Briefly, these protocols hide the items being requested even when servers collude with one another.
To this end, Sasidharan et al.\cite{sasidharan2021private} propose a Private Information Retrieval (PIR) protocol for sharded blockchain.  It allows checking of account balances assuming Reed-Solomon coded shards.
%

\section{Data Availability Attack}\label{sec_light}
Light nodes rely on full nodes to verify transactions are legitimate or correct~\cite{nakamoto2008bitcoin}.  Specifically, they are {\em not} responsible for checking whether a block producer's work is correct.  That responsibility lies with full nodes.  
In this respect, the data availability attack (DAA) is concerned with malicious full nodes that aim to convince light nodes to accept blocks with invalid or missing transactions~\cite{al2018fraud, battaglioni2022data}.    Specifically, a malicious full node can propagate a block with fraudulent transaction(s) to light nodes, which then readily accept the block as long as it is part of the longest blockchain.
A notable example was demonstrated by Peter Todd at the 2016 MIT Bitcoin Expo~\cite{Peter2016}, where he showed a light node that is fooled into thinking he owns 21 million bitcoins!
%
%
%

Addressing DAA involves a number of issues.  The first is data availability, meaning all data or transactions must be available to be checked by any nodes.  This discourages malicious behaviors as all data are public, and thus it can be verified by any parties.  In this respect, it is thus important to make data unavailability difficult.    Moreover, DAA can be detected by a node if a full node does not reply to (anonymous) requests for some transactions.
As we will discuss later, this problem can be addressed using erasure codes and random sampling.
Second, light nodes must incur low computational and communication cost in order to ascertain the validity of a block transmitted by a full node.  Further, a full node that generated a proof must provide supporting data that a transaction or coded symbol is invalid.
In this respect, exploiting Merkle proof is a key ingredient to such a method.
%
%
In summary, we need method(s) to prove fraud committed by nodes, aka {\em fraud proof}.  Further, we require a proof that all data is available, so called {\em data availability proof}. 
Further, the cost to download and verify a fraud proof must be low.  In addition, any developed schemes should not rely on having more honest full nodes than malicious full nodes.

%
%
\begin{table*}[htbp]
\begin{threeparttable}
    \centering
    \caption{Summary of DAA solutions for SPV (light) nodes.  The term $N$ is the number of nodes, $T$ is the number of blocks, $B$ is the block size, $k$ is the number of fragments to be encoded in a block.}
    \begin{tabular}{|m{2cm}|>{\centering}m{1.8cm}|>{\centering}m{1.8cm}|>{\centering}m{1cm}|>{\centering}m{1cm}|>{\centering}m{1.2cm}| m{6cm}|}
    \hline
          {\bf Reference} & {\bf Light node storage} & {\bf Light node communication}& {\bf Random  sample} & {\bf Require full node} & {\bf Code} & {\bf Key ideas}   \\
       \hline
        Al-Bassam et al. \cite{al2018fraud} & $\mathcal{O}(T\sqrt{B})$ & $\mathcal{O}(N\sqrt{B}\log B)$ & \checkmark    & \checkmark & RS & 
        \begin{tabenum}
            \item Encode block into coded fragments using RS code.
            \item Light nodes randomly request coded fragments from full nodes and send to a full node to decode.
            \item Hidden information is retrieved after decoding.
            \item Honest full nodes send fraud proof if there is inconsistency.
        \end{tabenum}
 \\
        \hline
         Santini et al.~\cite{santini2022optimization} & $\mathcal{O}(T\sqrt{B})$ & $\mathcal{O}(N\sqrt{B}\log B)$  & \checkmark  & \checkmark & RS  & 
         \begin{tabenum}
            \item Model the adversarial probability in \cite{al2018fraud} as a coupon collector's problem
            \item Optimize finite field size and communication cost given the number of light nodes
         \end{tabenum}
                  \\
        \hline
         SPAR \cite{yu2020coded} & $\mathcal{O}(T)$ & $\mathcal{O}(\log NB)$ & \checkmark    & \checkmark & LDPC & \begin{tabenum}
            \item Construct coded Merkle tree
            \item Reduce the number of coded fragment that light nodes need to transmit
            \item Reduce computational complexity for fraud proofs
         \end{tabenum}
         \\
        \hline
        Mitra et al. \cite{mitra2021overcoming} & $\mathcal{O}(T)$ & $\mathcal{O}(\log NB)$ & \checkmark  & \checkmark & LDPC &
        \begin{tabenum}
            \item Construct LDPC codes that reduce decoding failure probability
            \item Consider weak and strong adversary model
         \end{tabenum}
        \\
        \hline
        Mitra et al. \cite{mitra2022polar} & $\mathcal{O}(T)$ & $\mathcal{O}(\log NB)$ & \checkmark   & \checkmark & Polar & Construct coded merkle tree using Polar code. \\
        \hline
        Cao et al. \cite{cao2020cover} & $\mathcal{O}(T + \frac{1}{k}B\log B)$ & $\mathcal{O}(k \log N)$ & \checkmark   & X & LDPC &  
        \begin{tabenum}
            \item Light nodes issue fraud proof without full nodes
            \item Does not require locally majority of honest
         \end{tabenum}
        \\
        \hline
        
        
    \end{tabular}
    \label{tab_daa}
\end{threeparttable}
\end{table*}

\subsection{Simple Payment Verification (SPV)}\label{sec_CMT}
This section discusses approaches that aim to protect light nodes via fraud proofs.  A summary of these approaches is shown in Table \ref{tab_daa}. These approaches are designed to allow light nodes to identify/check a fraudulent block efficiently using limited amount of data.  The key concerns include {\em communication cost}, i.e., amount of data downloaded by light nodes to facilitate proofs checking, and {\em computation cost}, i.e., the run-time complexity of algorithms used to validate a proof.
The seminal work by Al-Bassam et al. \cite{al2018fraud} propose two types of fraud proofs: state transitions and data availability.
First, recall that a blockchain transitions to a new state after {\em all} transactions in a block are committed, and there is one Merkle root for each block.  Al-Bassam et al. propose {\em intermediate} states, which are created following the processing of some number of transactions.  Each intermediate state has a corresponding Merkle root.  If an honest full node  finds that the Merkle root of an intermediate state is incorrect, it can then inform light nodes.   These light nodes only need to confirm the finding of the full node, which can be carried out efficiently.  Specifically, the said full node only needs to send light nodes the Merkle root in question and some data.  Advantageously, this method only requires one honest full node to detect a fraudulent block and inform all light nodes.

Second, Al-Bassam et al. propose the following innovations to ensure data availability: (i) Reed-Solomon codes, and (ii) random sampling.
Specifically, full nodes encode each block using a Reed-Solomon code. Light nodes send anonymized requests for symbols of a block and its Merkle proof to different full nodes.  Light nodes then transmit received symbols to connected full nodes.   Once an honest full node has received sufficient number of symbols, it reconstructs the block and checks its validity.  If a block is deemed to be invalid, the full node sends a fraud proof to light nodes.  This causes the block to be rejected by light nodes.   To prevent a fraud proof from being issued, a malicious full node may not reply to requests from light nodes.  In this case, a light node ignores a block.  
Further, to prevent decoding of a block, malicious nodes may code a block incorrectly.  This is addressed using an incorrect decoding proof~\cite{yu2020coded}.
Note that applying only random sampling on un-coded or data blocks does not work.  This is because it is possible for a malicious full node to hide a small amount of data that cannot be detected via random sampling.  However, with an erasure code, missing data can be recovered after a node has received sufficient number of symbols.  This means a malicious node needs to hide more data to ensure a block is not decoded to be verified by a full node.  This, however, makes it easier to detect that some parts of a block are unavailable.
Apart from that, sampling requests must be anonymized to ensure a malicious full node does not target a light node selectively.
%
%

%
A key concern is the probability that a malicious node is able to deceive a light node when using the protocol in \cite{al2018fraud}; aka adversarial error probability. Another key concern is the signaling overheads generated by the random sampling process carried out by light nodes.
To this end, Santini et al.~\cite{santini2022optimization} propose a simple mathematical model to compute the said adversarial error probability. They relate the sampling process to the well-known coupon collector's problem~\cite{stadje1990collector}.  Advantageously, the model can be used to study the relationship between code rate, adversarial error probability, number of light nodes, and transmission overhead of light nodes.
Hence, it can be used to optimize finite field size and communication cost given the number of light nodes and desired adversarial error probability.
%
%
%

%
Communication cost incurred by light nodes to ensure data availability and to validate a block are of key interest to researchers.  
%
To this end, the authors of \cite{yu2020coded} propose SParse frAud pRotection (SPAR), 
which has a constant, i.e., $O(1)$, download cost (in bytes) for hash commitments (Merkle roots) and number of samples, $O(\sqrt{b} log(k))$ complexity for incorrect-coding proof and linear decoding complexity of $O(b)$.   Here $b$ is the block size, and $k$ is the number data symbols.  
The corresponding complexity for the work in~\cite{al2018fraud} is $O(\sqrt{b})$,
$O(1)$, $O(\sqrt{b} log(k))$ and $O(b^{1.5})$.
The key innovation in SPAR is a novel hash accumulator called Coded Merkle Tree (CMT).  It helps light nodes check the availability of coded symbols; note that CMT also helps a full node determine the membership of coded symbols, meaning every coded symbol for a given block has a Merkle proof.  
The key idea is to apply LDPC codes to each layer of a Merkle tree, where hashes are grouped into data symbols.  This in turn allows the application of the peeling decoder~\cite{2002LT} and reduces proof size, which is a function of the number of parity check equations~\cite{mitra2021communication}.
In addition, SPAR uses random sampling to ensure the CMT of a block is available.   Specifically, random sampling is used to check the availability of each CMT layer.
%
%

%
A key concern in~\cite{yu2020coded} is that the peeling decoder may fail if malicious nodes deliberately hide coded symbols in the stopping set of an LPDC code.  Specifically, the probability of decoding failure is a function of the smallest stopping set~\cite{LDPCStopDi}.
This induces a decoding failure whereby a full node is unable to determine the availability of a CMT.  This means full nodes are unable to generate an incorrect decoding proof or that there is DAA.
Thus, a key aim is to construct an LDPC code with a large minimum stopping set~\cite{mitra2021StopSet}.  This in turn allows light nodes, which use random sampling, to detect DAA with a high probability.  Unfortunately, constructing such a LDPC code is known to be an NP-hard problem~\cite{LDPCStopDi}.
To address this problem, Mitra et al. \cite{mitra2021StopSet} design specialized LDPC codes and a greedy sampling strategy.    Further, they note that the LDPC codes used in \cite{yu2020coded} are constructed using standard methods and optimized for the binary symmetric channel (BSC) as opposed to detecting DAA in blockchain systems.
To this end, the work in~\cite{mitra2021StopSet} builds on progressive edge-growth  (PEG) \cite{1023752} to construct LDPC codes that require light nodes to only sample a small number of variable nodes in order to detect whether a full node is hiding symbols in the stopping sets of an LDPC code. These set of variable nodes have the highest probability or entropy in being part of stopping sets.
In their follow up work~\cite{mitra2021overcoming}, they further consider weak and strong adversary models, respectively.  A weak attacker only interrupts decoding at one layer of the CMT, while a strong attacker can interrupt decoding of certain symbols at all layers.  Apart from that, for the strong adversary model, they outline a linear programming based method that takes as input all stopping sets with a given size for a given LDPC code; this exhaustive collection of stopping sets can be obtained for short code lengths using an integer linear program.
In a subsequent work, in \cite{mitra2022polar}, Mitra et al. construct a coded Merkle tree using Polar codes~\cite{PolarCodes}. 

%
Another approach to validate blocks is to bypass full nodes. To this end, Cao et al. \cite{cao2020cover} propose a collaborative verification protocol that involves only light nodes.   Each light node checks some transactions in a block and issues a fraud proof if necessary. Advantageously, their protocol scales sub-linearly with block size in terms of communication, computation and storage.  Further, it does not require a majority of honest light nodes.
Their protocol has a number of features.
First, they design a new block structure that allows light nodes to check each transaction in a block; this is similar to the work in~\cite{al2018fraud} but considers states that are distributed to light nodes.
Second, each light node verifies transactions belonging to specific senders/accounts.  Consequently, the approach in \cite{cao2020cover} is secure if all senders of transactions of a block are covered/verified by an honest light node.
Third, to ensure data availability, Cao et al. designed a protocol that uses the coded Merkle tree of~\cite{yu2020coded}.  Specifically, light nodes randomly sample symbols from each layer of the Merkle tree.
Further, to reduce storage requirement, each light node only processes transactions in a fixed number of blocks. 
Lastly, Cao et al. equip light nodes with a gossip protocol, e.g.,~\cite{GossipP2007}, to obtain symbols of interest and propagate fraud proofs.
%

%
%

%
\begin{figure}[t]
\centering
\subfigure[In conventional side chain solutions, each side chain submits a commitment, i.e., the hash of a block, to the main chain. The main chain then ensures the data integrity of side chains. However,  a malicious side chain node may submit a commitment without sharing it with others in the same side chain network. This lead to DAA. ]{%
\includegraphics[width=0.9\linewidth]{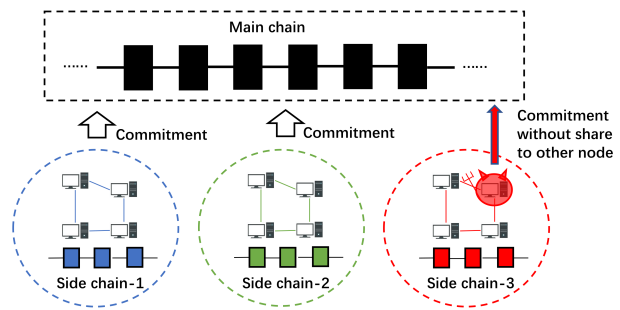}
\label{sidechain_old}}
\subfigure[By using an error correction code, a group of nodes form a committee and construct an oracle layer that interfaces side chains to the main chain. Side chain nodes submit coded blocks instead of a commitment to oracle layer nodes, which then decode and verify the content therein. If there is a decoding failure, oracle nodes send a fraud proof to inform other nodes in the corresponding side chain.   ]{%
\includegraphics[width=0.9\linewidth]{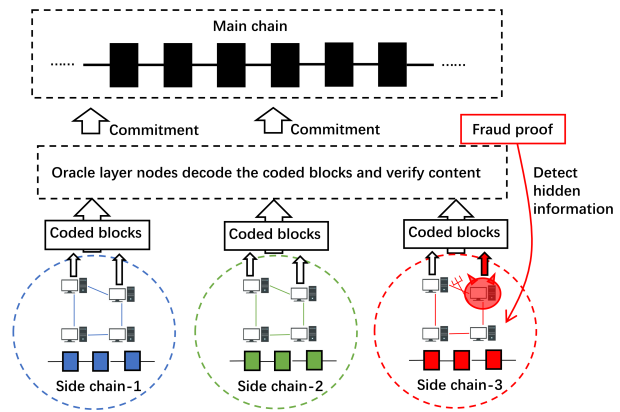}
\label{sidechain_coded}}

\caption{Comparison between conventional and coded side chains. }
\label{sidechains}
\end{figure}
%
\subsection{Side Chains}
Side chains aim to improve throughput and reduce the storage requirement of a blockchain system~\cite{BackPegged, SideChain2020}.  Further, it helps facilitate innovations, where the main features, e.g., consensus protocol, of the main chain remain intact.  A side chain, however, can incorporate new trust models and/or algorithms/protocols to process transactions~\cite{BackPegged}.
A key idea is the transfer of assets, e.g., cryptocurrencies, between the main chain and side chains~\cite{BackPegged}.  
For example, the authors of~\cite{SpritesPay} propose an off-chain payment protocol, where users use a main chain to setup a collateral or deposits and to settle disputes.  Advantageously, arbitrary number of payments are executed using a side chain or payment channel between users.  The main chain is only used to consolidate the account balance of users.  Another example is Arbitrum~\cite{217511}, which aims to scale smart contracts by moving the verification of their executions to a side chain or off-chain.


\begin{table*}[htbp]
    \centering
    \caption{Summary of coded blockchain solutions. }
    \begin{tabular}{|m{2cm}|>{\centering}m{4cm}| >{\centering\arraybackslash}m{10.5cm}|}
    \hline
          {\bf Aim} & {\bf Issues} & {\bf Solutions \& Reference}   \\
       \hline
         \multirow{2}{*}[-15pt]{Storage}     & Distribute storage of coded blockchain & 
         \begin{tabenum}
             \item Distribute storage use Network code  \cite{dai2018low}
             \item Operating with nodes storing coded blockchain\cite{perard2018erasure}\cite{wu2020distributed} \cite{qin2022downsampling}
         \end{tabenum}\\
           \cline{2-3}
          & Secure the distributed stored coded blockchain data & 
         \begin{tabenum}
             \item Secret key sharing \cite{raman2021ToN} \cite{mesnager2020threshold}
             \item Secret key sharing of coded block \cite{9771862}
             \item Global and local keys to reduce communication cost \cite{kim2019efficient}
             \item Adjust storage ratio in dynamic networks with at most a third are byzantine nodes~\cite{qi2020bft}
         \end{tabenum}\\
        \hline
          \multirow{3}{*}[-20pt]{Communication}    & Communication cost &  
          \begin{tabenum}
             \item Use network coding to reduce re-transmission cost \cite{cebe2018network}\cite{9609913}
             \item Reduce bandwidth requirement at bottleneck node \cite{chawla2019velocity}\cite{choi2019scalable}
             \item Reconstruct transactions and use code to restore missing transactions \cite{8922597}
             \item Use network coding to reduce message overhead in BFT consensus\cite{HBadger}\cite{BEATBFT}\cite{DLedgerTse}
         \end{tabenum} \\
        \cline{2-3}
          & Communication delay &  
          \begin{tabenum}
             \item Directly forward coded fragments \cite{zhang2022speeding}
             \item Optimize coded fragments storage and transmission \cite{li2021lightweight}\cite{yang2021storage}
             \item Locally encode and disperse coded fragments \cite{2020GCBlock}
         \end{tabenum} \\
        \cline{2-3}
            & Bootstrap new nodes &  
          \begin{tabenum}
             \item Reduce bandwidth requirement of full nodes \cite{kadhe2019sef}
             \item Ensure the correctness of block headers \cite{pal2020fountain}
             \item Reduce decoding complexity \cite{tiwari2021secure}
             \item New node directly stores coded fragment without decoding \cite{gadiraju2020secure}\cite{mitra2019patterned}
         \end{tabenum} \\
        \hline
          Computation    & Improve the security of sharded blockchains & 
          \begin{tabenum}
             \item Distribute computing among all shards \cite{polyshard2018}
             \item Enable cross shard transactions in coded shards \cite{wang2021low},\cite{wang2022JSIT}
             \item Differential attack for coded shards \cite{khooshemehr2021discrepancy}
             \item Ensure privacy of cross shards transactions \cite{sasidharan2021private}
             \item Reduce computation at each node \cite{asheralieva2021throughput}
         \end{tabenum}
         \\
        \hline
         \multirow{2}{*}[-3pt]{Security}    & Improve the security of SPV clients & 
         \begin{tabenum}
            \item Send a fraud alarm to a light node after a full node decodes a block \cite{al2018fraud}\cite{santini2022optimization}
            \item Generate a fraud alarm using a coded Merkle tree \cite{yu2020coded}\cite{cao2020cover} \cite{mitra2021StopSet}  \cite{mitra2021overcoming}\cite{mitra2022polar}\cite{battaglioni2022data}
         \end{tabenum}\\
        \cline{2-3}
          & Improve side chains security & 
        \begin{tabenum}
         \item Encode blocks, and have them decoded and checked by some oracle nodes\cite{sheng2020aced}\cite{mitra2021communication}
         \end{tabenum}
         \\
        \hline
     \end{tabular}
    \label{tab_summary}
\end{table*}


Side chains usually store the hash of blocks or executions states from side chains.  This means the main chain is used only to guarantee the data integrity of side chains.  For example, in Arbitrum~\cite{217511}, miners only need to verify hashes of smart contract states.
%
However, this may result in DAA because the main chain is not able to verify the said transactions or executions of a contract.  Hence, a key requirement is that a side chain node feeds the correct hash to the main chain.  Unfortunately, a malicious side chain node can feed the hash of a block to the main chain but not share the block~\cite{sheng2020aced}. 
To address the above issue, Sheng et al. \cite{sheng2020aced} propose to add an oracle layer between a side side and main chain.  Side chain nodes must first send LDPC coded blocks to nodes in the oracle layer. These so called oracle nodes then decode and verify the correctness of these blocks.  Only valid blocks are fed to the main chain. 
Mitra et al. \cite{mitra2021communication} build on their coded Merkle tree work \cite{mitra2021overcoming} to design an LDPC code for a side chain oracle. 
They focus on communication efficiency where oracle nodes verify blocks fed from side chain nodes.  They propose an algorithm based on PEG~\cite{1023752} that maximizes decoding probability using the fewest number of coded blocks. 
%

%
%
%
%

\section{Conclusion and Future works}
The growth in blockchain technologies and wide ranging applications have resulted in scaling issues relating to data storage, communication, computation and security.
To this end, this paper considers approaches that employ error correction codes.   Some of their main advantages include (i) reduced storage and communication cost, and thus allowing more nodes, especially light nodes, to participate in a blockchain.  Further, communication cost is reduced as as each node is only required to communicate a small proportion of coded fragments.  In addition, it facilitates bootstrapping of new nodes, (ii) better security, where coded blockchains are more resistant to manipulation by malicious nodes and promote data availability.
%
%
These key features are highlighted in Table~\ref{tab_summary}.

%
There are a number of possible future works.
To date, no prior coded blockchain approaches have used machine learning approaches to optimize coding parameters for different network conditions.
For example, the work in \cite{5158447} shows that computational cost can be significant if parameters of regenerating codes are not chosen correctly.  Another example is to identify patterned erasures due to nodes leaving a blockchain system caused by failures or duty cycling in order to conserve energy.  
Another avenue is to use machine learning approaches to determine the degree distribution of fountain codes such as LT, e.g., \cite{8845185}.
Further, coding parameters must be adjusted subject to the cost and availability of communication, computation and storage resources.

Another possible research direction relates to communication cost.
A number of works use Reed-Solomon codes, e.g.,~\cite{li2021lightweight, qi2020bft}.  However, maximum distance separable codes are known to have a large repair bandwidth~\cite{XORE13}.  This has implication on light nodes that have limited bandwidth.
Further, for works that use network coding, e.g.,~\cite{9609913}, the number of coefficients used to encode fragments have an impact on communication cost.  A key challenge is the dynamic topology of a blockchain system.  Hence, adapting coefficients over time whilst ensuring low communication cost and decoding success is critical.
Apart from that, there are limited number of works that consider the underlying network conditions when distributing or coding blocks.  For example, participants may be far apart, which increases propagation delays, or connected via congested links, which delays consensus time and distribution of coded fragments.

\bibliographystyle{ieeetr}
\bibliography{ref}

\begin{thebibliography}{100}

\bibitem{underwood2016blockchain}
S.~Underwood, ``Blockchain beyond bitcoin,'' {\em Communications of the ACM},
  vol.~59, no.~11, pp.~15--17, 2016.

\bibitem{nakamoto2008bitcoin}
S.~Nakamoto, ``Bitcoin: A peer-to-peer electronic cash system,'' {\em
  Decentralized Business Review}, p.~21260, 2008.

\bibitem{ethereum}
``Ethereum,'' 2008.
\newblock [Online]. Available: \url{https://ethereum.org/en/}.

\bibitem{mettler2016blockchain}
M.~Mettler, ``Blockchain technology in healthcare: The revolution starts
  here,'' in {\em IEEE 18th international conference on e-health networking,
  applications and services (Healthcom)}, (Munich, Germany), pp.~1--3, Sept.
  2016.

\bibitem{BCsupply19}
Q.~Zhu and M.~Kouhizadeh, ``Blockchain technology, supply chain information,
  and strategic product deletion management,'' {\em IEEE Engineering Management
  Review}, vol.~47, no.~1, pp.~36 -- 44, 2019.

\bibitem{MaDRBC}
Z.~Ma, M.~Jiang, H.~Gao, and Z.~Wang, ``Blockchain for digital rights
  management,'' {\em Future Generation Computer Systems}, vol.~89,
  pp.~746--764, Dec. 2018.

\bibitem{chen2017improved}
Y.~Chen, H.~Li, K.~Li, and J.~Zhang, ``An improved {P2P} file system scheme
  based on {IPFS} and blockchain,'' in {\em IEEE International Conference on
  Big Data (Big Data)}, (Boston, USA), pp.~2652--2657, Dec. 2017.

\bibitem{teslya2017blockchain}
N.~Teslya and I.~Ryabchikov, ``Blockchain-based platform architecture for
  industrial {IoT},'' in {\em 21st Conference of Open Innovations Association
  (FRUCT)}, (Helskin, Finland), pp.~321--329, Nov. 2017.

\bibitem{BCDomains}
w.~Chen, Z.~Xu, S.~Shi, Y.~Zhao, and J.~Zhao, ``A survey of blockchain
  applications in different domains,'' in {\em International Conference on
  Blockchain Technology and Application}, (Xian, China), pp.~1--6, Dec. 2018.

\bibitem{castro1999practical}
M.~Castro and B.~Liskov, ``Practical byzantine fault tolerance and proactive
  recovery,'' in {\em Operating Systems Design and Implementation}, (New
  Orleans, USA), pp.~173--186, 1999.

\bibitem{nakamoto2008peer}
``A peer-to-peer electronic cash system,'' 2008.
\newblock [Online]. Available: \url{https://bitcoin. org/bitcoin. pdf}.

\bibitem{bitcoinsize}
``Blockchain size,'' 2021.
\newblock [Online]. Available:
  \url{https://www.blockchain.com/charts/blocks-size}.

\bibitem{Ripple2019}
``Capacity planning.''
\newblock \url{https://developers.ripple.com/capacity-planning.html},[Online;
  Accessed on 06/20/2019].

\bibitem{ripple_size}
``{XRP} ledger, ripple blockchain size,'' 2008.
\newblock [Online]. Available: \url{https://xrpl.org/capacity-planning.html}.

\bibitem{raman2021ToN}
R.~K. Raman and L.~R. Varshney, ``Coding for scalable blockchains via dynamic
  distributed storage,'' {\em IEEE/ACM Trans. on Netw.}, vol.~29,
  pp.~2588--2601, Dec. 2021.

\bibitem{nadiya2018block}
U.~Nadiya, K.~Mutijarsa, and C.~Y. Rizqi, ``Block summarization and compression
  in bitcoin blockchain,'' in {\em International Symposium on Electronics and
  Smart Devices (ISESD)}, (Indonesia), pp.~1--6, Oct. 2018.

\bibitem{kim2019scc}
T.~Kim, J.~Noh, and S.~Cho, ``{SCC}: Storage compression consensus for
  blockchain in lightweight {IoT} network,'' in {\em IEEE International
  Conference on Consumer Electronics (ICCE)}, (Las Vegas, USA), pp.~1--4, Jan.
  2019.

\bibitem{Mina}
``Mina protocol,'' 2008.
\newblock [Online]. Available: \url{https://minaprotocol.com/}.

\bibitem{zamani2018rapidchain}
M.~Zamani, M.~Movahedi, and M.~Raykova, ``{Rapidchain:} scaling blockchain via
  full sharding,'' in {\em ACM SIGSAC Conference on Computer and Communications
  Security}, (Toronto, Canada), pp.~931--948, Oct. 2018.

\bibitem{karame2016bitcoin}
G.~O. Karame and E.~Androulaki, {\em Bitcoin and Blockchain Security}.
\newblock Artech House, 2016.

\bibitem{perard2018erasure}
D.~Perard, J.~Lacan, Y.~Bachy, and J.~Detchart, ``Erasure code-based low
  storage blockchain node,'' in {\em IEEE International Conference on Internet
  of Things (iThings) and IEEE Green Computing and Communications (GreenCom)
  and IEEE Cyber, Physical and Social Computing (CPSCom) and IEEE Smart Data
  (SmartData)}, (Halifax, NS, Canada), pp.~1622--1627, July 2018.

\bibitem{das2018security}
S.~Das, A.~Kolluri, P.~Saxena, and H.~Yu, ``On the security of blockchain
  consensus protocols,'' in {\em International Conference on Information
  Systems Security}, (Funchal, Portugal), pp.~465--480, Jan. 2018.

\bibitem{Dimakis2011Survey}
A.~G. Dimakis, K.~Ramchandran, Y.~Wu, and C.~Suh, ``A survey on network codes
  for distributed storage,'' {\em Proc. IEEE}, vol.~99, no.~3, pp.~476--489,
  2011.

\bibitem{CachinIDV}
C.~Cachin and S.~Tessaro, ``Asynchronous verifiable information dispersal,'' in
  {\em 24th IEEE Symposium on Reliable Distributed Systems}, (Orlando, FL,
  USA), pp.~1--11, Oct. 2005.

\bibitem{1994Reed}
S.~B. Wicker and V.~K. Bhargava, {\em Reed-Solomon codes and their
  applications}.
\newblock Wiley-IEEE Press, 1994.

\bibitem{Ethereum_light_node}
``Ethereum light node,'' 2008.
\newblock [Online]. Available:
  \url{https://ethereum.org/en/developers/tutorials/run-light-node-geth/}.

\bibitem{pruned_node_space}
``Bitcoin pruned node,'' 2008.
\newblock [Online]. Available:
  \url{https://bitcoin.org/en/full-node#reduce-storage}.

\bibitem{dai2018low}
M.~Dai, S.~Zhang, H.~Wang, and S.~Jin, ``A low storage room requirement
  framework for distributed ledger in blockchain,'' {\em IEEE Access}, vol.~6,
  pp.~22970--22975, 2018.

\bibitem{wu2020distributed}
H.~Wu, A.~Ashikhmin, X.~Wang, C.~Li, S.~Yang, and L.~Zhang, ``Distributed error
  correction coding scheme for low storage blockchain systems,'' {\em IEEE
  Internet Things J.}, vol.~7, no.~8, pp.~7054--7071, 2020.

\bibitem{quan2019transparent}
L.~Quan and Q.~Huang, ``Transparent coded blockchain,'' in {\em ACM CoNEXT},
  (Orlando, USA), pp.~12--13, Dec. 2019.

\bibitem{li2021lightweight}
C.~Li, J.~Zhang, X.~Yang, and L.~Youlong, ``Lightweight blockchain consensus
  mechanism and storage optimization for resource-constrained {IoT} devices,''
  {\em Information Processing \& Management}, vol.~58, no.~4, p.~102602, 2021.

\bibitem{qin2022downsampling}
Q.~Huang, L.~Quan, and S.~Zhang, ``Downsampling and transparent coding for
  blockchain,'' {\em IEEE Trans. Netw. Sci. Eng.}, vol.~9, pp.~2139 -- 2149,
  Aug. 2022.

\bibitem{raman2017dynamic}
R.~K. Raman and L.~R. Varshney, ``Dynamic distributed storage for scaling
  blockchains.'' arXiv preprint arXiv:1711.07617, 2017.

\bibitem{raman2018distributed}
R.~K. Raman and L.~R. Varshney, ``Distributed storage meets secret sharing on
  the blockchain,'' in {\em Information Theory and Applications Workshop
  (ITA)}, (San Diego, CA, USA), pp.~1--6, Feb. 2018.

\bibitem{raman2018dynamic}
R.~K. Raman and L.~R. Varshney, ``Dynamic distributed storage for
  blockchains,'' in {\em IEEE International Symposium on Information Theory
  (ISIT)}, (Colorado, USA), pp.~2619--2623, June 2018.

\bibitem{kim2019efficient}
Y.~Kim, R.~K. Raman, Y.-S. Kim, L.~R. Varshney, and N.~R. Shanbhag, ``Efficient
  local secret sharing for distributed blockchain systems,'' {\em IEEE Commun.
  Letters}, vol.~23, no.~2, pp.~282--285, 2019.

\bibitem{mesnager2020threshold}
S.~Mesnager, A.~S{\i}nak, and O.~Yayla, ``Threshold-based post-quantum secure
  verifiable multi-secret sharing for distributed storage blockchain,'' {\em
  Mathematics}, vol.~8, no.~12, p.~2218, 2020.

\bibitem{qi2020bft}
X.~Qi, Z.~Zhang, C.~Jin, and A.~Zhou, ``{BFT-Store:} storage partition for
  permissioned blockchain via erasure coding,'' in {\em IEEE 36th International
  Conference on Data Engineering (ICDE)}, (Dallas, USA), pp.~1926--1929, Apr.
  2020.

\bibitem{9771862}
J.~Singh, A.~Banerjee, and H.~Sadjadpour, ``Secure and private fountain code
  based architecture for blockchains,'' in {\em IEEE WCNC}, (Austin, TX, USA),
  pp.~1521--1526, Apr. 2022.

\bibitem{HBadger}
A.~Miller, Y.~Xia, and K.~Croman, ``The honey badger of {BFT} protocols,'' in
  {\em Proceedings of the ACM SIGSAC Conference on Computer and Communications
  Security}, (Vienna, Austria), pp.~1--14, Oct. 2016.

\bibitem{cebe2018network}
M.~Cebe, B.~Kaplan, and K.~Akkaya, ``A network coding based information
  spreading approach for permissioned blockchain in {IoT} settings,'' in {\em
  ACM Mobiquitous}, (New York, USA), pp.~470--475, Nov. 2018.

\bibitem{BEATBFT}
S.~Duan, M.~K. Reiter, and H.~Zhang, ``{BEAT:} asynchronous {BFT} made
  practical,'' in {\em Proceedings of the ACM SIGSAC Conference on Computer and
  Communications Security}, (Toronto, Canada), pp.~1--14, Oct. 2018.

\bibitem{choi2019scalable}
B.~Choi, J.-y. Sohn, D.-J. Han, and J.~Moon, ``Scalable network-coded {PBFT}
  consensus algorithm,'' in {\em IEEE International Symposium on Information
  Theory (ISIT)}, (Paris, France), pp.~857--861, July 2019.

\bibitem{chawla2019velocity}
N.~Chawla, H.~W. Behrens, D.~Tapp, D.~Boscovic, and K.~S. Candan, ``Velocity:
  Scalability improvements in block propagation through rateless erasure
  coding,'' in {\em IEEE International Conference on Blockchain and
  Cryptocurrency (ICBC)}, (Seoul, South Korea), pp.~447--454, May 2019.

\bibitem{8922597}
M.~Jin, X.~Chen, and S.-J. Lin, ``Reducing the bandwidth of block propagation
  in bitcoin network with erasure coding,'' {\em IEEE Access}, vol.~7,
  pp.~175606--175613, Dec. 2019.

\bibitem{2020GCBlock}
B.~Qu, L.~E. Wang, P.~Liu, Z.~Shi, and X.~X. Li, ``{GCBlock:} a grouping and
  coding based storage scheme for blockchain system,'' {\em IEEE Access},
  vol.~PP, no.~99, pp.~1--1, 2020.

\bibitem{9609913}
M.~Braun, A.~Wiesmaier, N.~Alnahawi, and J.~{\ss}eibler, ``On message-based
  consensus and network coding,'' in {\em 12th International Conference on
  Network of the Future (NoF)}, (Coimbra, Portugal), pp.~1--9, 2021.

\bibitem{yang2021storage}
C.~Yang, X.~Wang, and A.~Ashikhmin, ``Storage and communication tradeoff for
  wireless coded blockchains,'' {\em IEEE Systems Journal}, vol.~16, pp.~2911
  -- 2922, June 2022.

\bibitem{DLedgerTse}
L.~Yang, S.~J. Park, M.~Alizadeh, S.~Kannan, and D.~Tse, ``{DispersedLedger}:
  High-throughput byzantine consensus on variable bandwidth networks,'' in {\em
  USENIX NSDI}, (Renton, WA, USA), pp.~493--512, Oct. 2022.

\bibitem{zhang2022speeding}
L.~Zhang, T.~Wang, and S.~C. Liew, ``Speeding up block propagation in bitcoin
  network: Uncoded and coded designs,'' {\em Computer Networks}, vol.~206,
  p.~108791, Apr. 2022.

\bibitem{kadhe2019sef}
S.~Kadhe, J.~Chung, and K.~Ramchandran, ``{SeF}: A secure fountain architecture
  for slashing storage costs in blockchains.'' arXiv:1906.12140, Jan. 2019.

\bibitem{mitra2019patterned}
D.~Mitra and L.~Dolecek, ``Patterned erasure correcting codes for low
  storage-overhead blockchain systems,'' in {\em 53rd Asilomar Conference on
  Signals, Systems, and Computers}, (Pacific Grove, CA, USA), pp.~1734--1738,
  Nov. 2019.

\bibitem{pal2020fountain}
R.~Pal, ``Fountain coding for bootstrapping of the blockchain,'' in {\em IEEE
  COMSNETS}, (Bengaluru, India), pp.~1--5, Jan. 2020.

\bibitem{gadiraju2020secure}
D.~S. Gadiraju, V.~Lalitha, and V.~Aggarwal, ``Secure regenerating codes for
  reducing storage and bootstrap costs in sharded blockchains,'' in {\em IEEE
  International Conference on Blockchain}, (Rhodes Island, USA), pp.~229--236,
  Nov. 2020.

\bibitem{tiwari2021secure}
A.~Tiwari and V.~Lalitha, ``Secure raptor encoder and decoder for low storage
  blockchain,'' in {\em IEEE COMSNET}, (Bangalore, India), pp.~161--165, Jan.
  2021.

\bibitem{polyshard2018}
S.~Li, M.~Yu, C.-S. Yang, A.~S. Avestimehr, S.~Kannan, and P.~Viswanath,
  ``{PolyShard}: Coded sharding achieves linearly scaling efficiency and
  security simultaneously.'' arXiv preprint arXiv:1809.10361, 2018.

\bibitem{2020PolyShard}
S.~Li, M.~Yu, C.~S. Yang, A.~S. Avestimehr, and P.~Viswanath, ``Polyshard:
  Coded sharding achieves linearly scaling efficiency and security
  simultaneously,'' {\em IEEE Trans. Inf. Forensics Security}, vol.~16, pp.~249
  -- 261, July 2020.

\bibitem{wang2021low}
C.~Wang and N.~Raviv, ``Low latency cross-shard transactions in coded
  blockchain,'' in {\em IEEE International Symposium on Information Theory
  (ISIT)}, (Melbourne, Australia), pp.~2678--2683, July 2021.

\bibitem{khooshemehr2021discrepancy}
N.~A. Khooshemehr and M.~A. Maddah-Ali, ``The discrepancy attack on
  polyshard-ed blockchains,'' in {\em IEEE International Symposium on
  Information Theory (ISIT)}, (Melbourne, Australia), pp.~2672--2677, July
  2021.

\bibitem{sasidharan2021private}
B.~Sasidharan and E.~Viterbo, ``Private data access in blockchain systems
  employing coded sharding,'' in {\em IEEE International Symposium on
  Information Theory (ISIT)}, (Melbourne, Australia), pp.~2684--2689, July
  2021.

\bibitem{asheralieva2021throughput}
A.~Asheralieva and D.~Niyato, ``Throughput-efficient lagrange coded private
  blockchain for secured {IoT} systems,'' {\em IEEE Internet Things J.},
  vol.~8, no.~19, pp.~14874--14895, 2021.

\bibitem{wang2022JSIT}
C.~Wang and N.~Raviv, ``Breaking blockchain's communication barrier with coded
  computation,'' {\em IEEE Journal on Selected Areas in Information Theory},
  July 2022.

\bibitem{al2018fraud}
M.~Al-Bassam, A.~Sonnino, and V.~Buterin, ``Fraud and data availability proofs:
  Maximising light client security and scaling blockchains with dishonest
  majorities.'' arXiv preprint arXiv:1809.09044, 2018.

\bibitem{yu2020coded}
M.~Yu, S.~Sahraei, S.~Li, S.~Avestimehr, S.~Kannan, and P.~Viswanath, ``Coded
  merkle tree: Solving data availability attacks in blockchains,'' in {\em
  International Conference on Financial Cryptography and Data Security}, (Kota
  Kinabalu, Malaysia), pp.~114--134, Feb. 2020.

\bibitem{cao2020cover}
S.~Cao, S.~Kadhe, and K.~Ramchandran, ``{CoVer}: Collaborative light-node-only
  verification and data availability for blockchains,'' in {\em IEEE
  International Conference on Blockchain}, (Rhodes Island, USA), pp.~45--52,
  Nov. 2020.

\bibitem{sheng2020aced}
P.~Sheng, B.~Xue, S.~Kannan, and P.~Viswanath, ``{ACeD:} scalable data
  availability oracle.'' arXiv preprint arXiv:2011.00102, 2020.

\bibitem{mitra2021StopSet}
D.~Mitra, L.~Tauz, and L.~Dolecek, ``Concentrated stopping set design for coded
  merkle tree: Improving security against data availability attacks in
  blockchain systems,'' in {\em IEEE Information Theory Workshop (ITW)}, (Riva
  del Garda), pp.~1--6, Apr. 2021.

\bibitem{mitra2021overcoming}
D.~Mitra, L.~Tauz, and L.~Dolecek, ``Overcoming data availability attacks in
  blockchain systems: {LDPC} code design for coded merkle tree.'' arXiv
  preprint arXiv:2108.13332, 2021.

\bibitem{mitra2021communication}
D.~Mitra, L.~Tauz, and L.~Dolecek, ``Communication-efficient {LDPC} code design
  for data availability oracle in side blockchains,'' in {\em IEEE Information
  Theory Workshop (ITW)}, (Kanazawa, Japan), pp.~1--6, Oct. 2021.

\bibitem{santini2022optimization}
P.~Santini, G.~Rafaiani, M.~Battaglioni, F.~Chiaraluce, and M.~Baldi,
  ``Optimization of a reed-solomon code-based protocol against blockchain data
  availability attacks.'' arXiv preprint arXiv:2201.08261, 2022.

\bibitem{mitra2022polar}
D.~Mitra, L.~Tauz, and L.~Dolecek, ``Polar coded merkle tree: Improved
  detection of data availability attacks in blockchain systems,'' in {\em IEEE
  International Symposium on Information Theory (ISIT)}, (Espoo, Finland), June
  2022.

\bibitem{battaglioni2022data}
M.~Battaglioni, P.~Santini, G.~Rafaiani, F.~Chiaraluce, and M.~Baldi, ``A data
  availability attack on a blockchain protocol based on {LDPC} codes.'' arXiv
  preprint arXiv:2202.07265, 2022.

\bibitem{zhou2020solutions}
Q.~Zhou, H.~Huang, Z.~Zheng, and J.~Bian, ``Solutions to scalability of
  blockchain: A survey,'' {\em IEEE Access}, vol.~8, pp.~16440--16455, 2020.

\bibitem{kim2018survey}
S.~Kim, Y.~Kwon, and S.~Cho, ``A survey of scalability solutions on
  blockchain,'' in {\em International Conference on Information and
  Communication Technology Convergence (ICTC)}, (Jeju Island , South Korea),
  pp.~1204--1207, Oct. 2018.

\bibitem{8823874}
J.~Xie, F.~R. Yu, T.~Huang, R.~Xie, J.~Liu, and Y.~Liu, ``A survey on the
  scalability of blockchain systems,'' {\em IEEE Network}, vol.~33, no.~5,
  pp.~166--173, 2019.

\bibitem{hafid2020scaling}
A.~Hafid, A.~S. Hafid, and M.~Samih, ``Scaling blockchains: A comprehensive
  survey,'' {\em IEEE Access}, vol.~8, pp.~125244--125262, 2020.

\bibitem{yu2020survey}
G.~Yu, X.~Wang, K.~Yu, W.~Ni, J.~A. Zhang, and R.~P. Liu, ``Survey: Sharding in
  blockchains,'' {\em IEEE Access}, vol.~8, pp.~14155--14181, 2020.

\bibitem{mazlan2020scalability}
A.~A. Mazlan, S.~M. Daud, S.~M. Sam, H.~Abas, S.~Z.~A. Rasid, and M.~F. Yusof,
  ``Scalability challenges in healthcare blockchain system—a systematic
  review,'' {\em IEEE Access}, vol.~8, pp.~23663--23673, 2020.

\bibitem{SideChain2020}
A.~Singh, K.~click, R.~M. Parizi, Q.~Zhang, A.~Dehghantanha, and K.~K.~R. Choo,
  ``Sidechain technologies in blockchain networks: An examination and
  state-of-the-art review,'' {\em Journal of Network and Computer
  Applications}, vol.~149, pp.~1--16, Jan. 2020.

\bibitem{sanka2021systematic}
A.~I. Sanka and R.~C. Cheung, ``A systematic review of blockchain scalability:
  Issues, solutions, analysis and future research,'' {\em Journal of Network
  and Computer Applications}, vol.~195, p.~103232, 2021.

\bibitem{nasir2022scalable}
M.~H. Nasir, J.~Arshad, M.~M. Khan, M.~Fatima, K.~Salah, and R.~Jayaraman,
  ``Scalable blockchains—a systematic review,'' {\em Future Generation
  Computer Systems}, vol.~126, pp.~136--162, 2022.

\bibitem{5550492}
A.~G. Dimakis, P.~B. Godfrey, Y.~Wu, M.~J. Wainwright, and K.~Ramchandran,
  ``Network coding for distributed storage systems,'' {\em IEEE Trans. Inf.
  Theory}, vol.~56, no.~9, pp.~4539--4551, 2010.

\bibitem{2002Low}
G.~Cancellieri, ``Low-density parity-check codes,'' {\em Wiley-IEEE Press},
  2002.

\bibitem{2002LT}
M.~Luby, ``{LT} codes,'' in {\em 43rd Symposium on Foundations of Computer
  Science (FOCS)}, (Vancouver, Canada), Nov. 2002.

\bibitem{2006Raptor}
A.~Shokrollahi, ``Raptor codes,'' {\em IEEE Trans. Inf. Theory}, vol.~52,
  no.~6, pp.~2551--2567, 2006.

\bibitem{RabinIDA}
M.~O. Rabin, ``Efficient dispersal of information for security, load balancing,
  and fault tolerance,'' {\em Journal of the ACM}, vol.~36, no.~2,
  pp.~335--348, 1989.

\bibitem{9445631}
R.~Zhang, R.~Xue, and L.~Liu, ``Security and privacy for healthcare
  blockchains,'' {\em IEEE Trans. Services Comput}, pp.~1--1, 2021.

\bibitem{Lock1}
Y.~Wang and A.~Kogan, ``Designing confidentiality-preserving blockchain-based
  transaction processing systems,'' {\em International Journal of Accounting
  Information Systems}, vol.~30, pp.~1--18, Sept. 2018.

\bibitem{Lock2Survey}
R.~Zhang, R.~Xue, and L.~Liu, ``Security and privacy on blockchain,'' {\em ACM
  Computing Surveys}, vol.~52, pp.~1--34, May 2020.

\bibitem{9093015}
D.~Wang, J.~Zhao, and Y.~Wang, ``A survey on privacy protection of blockchain:
  The technology and application,'' {\em IEEE Access}, vol.~8,
  pp.~108766--108781, May 2020.

\bibitem{8957108}
W.~Liang, Y.~Fan, K.-C. Li, D.~Zhang, and J.-L. Gaudiot, ``Secure data storage
  and recovery in industrial blockchain network environments,'' {\em IEEE
  Trans. Ind. Informat.}, vol.~16, pp.~6543--6552, Oct. 2020.

\bibitem{Shamir1979how}
A.~Shamir, ``How to share a secret,'' {\em Commun. ACM}, vol.~22, p.~612–613,
  nov 1979.

\bibitem{9174266}
R.~Bitar and S.~Jaggi, ``Communication efficient secret sharing in the presence
  of malicious adversary,'' in {\em IEEE International Symposium on Information
  Theory (ISIT)}, (Los Angeles, CA, USA), pp.~548--553, June 2020.

\bibitem{8006842}
W.~Huang and J.~Bruck, ``Secret sharing with optimal decoding and repair
  bandwidth,'' in {\em IEEE International Symposium on Information Theory
  (ISIT)}, (Aachen, Germany), pp.~1813--1817, June 2017.

\bibitem{feldman1987practical}
P.~Feldman, ``A practical scheme for non-interactive verifiable secret
  sharing,'' in {\em 28th Annual Symposium on Foundations of Computer Science},
  (Los Angeles, CA, USA), pp.~427--438, Oct. 1987.

\bibitem{9539193}
N.~Loizou and P.~Richt\'{a}rik, ``Revisiting randomized gossip algorithms:
  General framework, convergence rates and novel block and accelerated
  protocols,'' {\em IEEE Trans. Inf. Theory}, vol.~67, no.~12, pp.~8300--8324,
  2021.

\bibitem{miguel1999pbft}
M.~Castro and B.~Liskov, ``Practical byzantine fault tolerance and proactive
  recovery,'' {\em ACM Trans. Comput. Syst.}, vol.~20, p.~398–461, Nov. 2002.

\bibitem{6688704}
C.~Decker and R.~Wattenhofer, ``Information propagation in the bitcoin
  network,'' in {\em IEEE International Conference on Peer-to-Peer Computing},
  (Trento, Italy), pp.~1--10, Sept. 2013.

\bibitem{Gobel2017Increased}
J.~G\"{o}bel and A.~Krzesinski, ``Increased block size and bitcoin blockchain
  dynamics,'' in {\em 27th International Telecommunication Networks and
  Applications Conference (ITNAC)}, (Melbourne, Australia), Nov. 2017.

\bibitem{5061931}
J.~K. Sundararajan, D.~Shah, M.~Medard, M.~Mitzenmacher, and J.~Barros,
  ``Network coding meets {TCP},'' in {\em IEEE INFOCOM}, (Rio de Janeiro,
  Brazil), pp.~280--288, Apr. 2009.

\bibitem{5462030}
Z.~Liu, C.~Wu, B.~Li, and S.~Zhao, ``{UUSee}: Large-scale operational on-demand
  streaming with random network coding,'' in {\em IEEE INFOCOM}, (San Diego,
  CA, USA), pp.~1--9, Mar. 2010.

\bibitem{Bracha84}
G.~Bracha, ``An asynchronous $[(n-1)/3]$-resilient consensus protocol,'' in
  {\em Proceedings of the Third Annual ACM Symposium on Principles of
  Distributed Computing}, (Vancouver, Canada), pp.~154--162, Aug. 1984.

\bibitem{luu2016secure}
L.~Luu, V.~Narayanan, C.~Zheng, K.~Baweja, S.~Gilbert, and P.~Saxena, ``A
  secure sharding protocol for open blockchains,'' in {\em ACM SIGSAC
  Conference on Computer and Communications Security}, (Vienna, Austria),
  pp.~17--30, Oct. 2016.

\bibitem{ahlswede2000network}
R.~Ahlswede, N.~Cai, S.-Y. Li, and R.~W. Yeung, ``Network information flow,''
  {\em IEEE Trans. Inf. Theory}, vol.~46, no.~4, pp.~1204--1216, 2000.

\bibitem{9219639}
R.~Nagayama, R.~Banno, and K.~Shudo, ``Identifying impacts of protocol and
  internet development on the bitcoin network,'' in {\em IEEE Symposium on
  Computers and Communications (ISCC)}, (Rennes, France), July 2020.

\bibitem{Rouayheb2010Fractional}
S.~El~Rouayheb and K.~Ramchandran, ``Fractional repetition codes for repair in
  distributed storage systems,'' in {\em 48th Annual Allerton Conference on
  Communication, Control, and Computing (Allerton)}, (Chicago, USA),
  pp.~1510--1517, Sept. 2010.

\bibitem{kalman1984generalized}
D.~Kalman, ``The generalized vandermonde matrix,'' {\em Mathematics Magazine},
  vol.~57, no.~1, pp.~15--21, 1984.

\bibitem{SQLBook}
B.~Schwartz, P.~Zaitsev, V.~Tkachenko, J.~D. Zawodny, A.~Lentz, and D.~J.
  Balling, {\em High Performance {MySQL}: Optimization, Backups, Replication
  and More}.
\newblock O'Reilly Media, June 2008.

\bibitem{yu2019lagrange}
Q.~Yu, S.~Li, N.~Raviv, S.~M.~M. Kalan, M.~Soltanolkotabi, and S.~A.
  Avestimehr, ``Lagrange coded computing: Optimal design for resiliency,
  security, and privacy,'' in {\em The 22nd International Conference on
  Artificial Intelligence and Statistics}, pp.~1215--1225, 2019.

\bibitem{PQuery2001}
A.~Beimel and Y.~Ishai, ``Information-theoretic private information retrieval:
  A unified construction,'' in {\em International Colloquium on Automata,
  Languages, and Programming (ICALP)}, (Malaga, Spain), pp.~1--15, July 2001.

\bibitem{Peter2016}
P.~Todd.
  https://diyhpl.us/wiki/transcripts/mit-bitcoin-expo-2016/fraud-proofs-petertodd/,
  2016.

\bibitem{stadje1990collector}
W.~Stadje, ``The collector's problem with group drawings,'' {\em Advances in
  Applied Probability}, vol.~22, no.~4, pp.~866--882, 1990.

\bibitem{LDPCStopDi}
C.~Di, D.~Prioietti, E.~Telatar, T.~J. Richardson, and R.~L. Urbanke,
  ``Finite-length analysis of low-density parity-check codes on the binary
  erasure channel,'' {\em IEEE Trans. Inform. Theory}, vol.~48, pp.~1570--1580,
  June 2002.

\bibitem{1023752}
X.-Y. Hu, E.~Eleftheriou, and D.-M. Arnold, ``Irregular progressive edge-growth
  {(PEG)} tanner graphs,'' in {\em Proceedings IEEE International Symposium on
  Information Theory,}, (Lausanne, Switzerland), pp.~480--486, 2002.

\bibitem{PolarCodes}
P.~Trifonov, ``Efficient design and decoding of polar codes,'' {\em IEEE Trans.
  Commun.}, vol.~60, no.~11, pp.~3221--3227, 2012.

\bibitem{GossipP2007}
K.~Birman, ``The promise, and limitations, of gossip protocols,'' {\em ACM
  SIGOPS Operating Systems Review}, vol.~41, no.~5, pp.~8--13, 2007.

\bibitem{BackPegged}
A.~Back, M.~Corallo, L.~Dashjr, M.~Friedenbach, G.~Maxwell, A.~Miller,
  A.~Poelstra, J.~Tim\'{o}n, and P.~Wuille, ``Enabling blockchain innovations
  with pegged sidechains.'' \url{http://www.blockstream.com/sidechains.pdf}.
\newblock Accessed: 2022-07-26.

\bibitem{SpritesPay}
M.~Diouf, D.~Declercq, S.~Ouya, and B.~Vasic, ``Sprites and state channels:
  Payment networks that go faster than lightning,'' in {\em International
  Conference on Financial Cryptography and Data Security}, (Frigate Bay, Saint
  Kitts), p.~508–526, Feb. 2019.

\bibitem{217511}
H.~Kalodner, S.~Goldfeder, X.~Chen, S.~M. Weinberg, and E.~W. Felten,
  ``Arbitrum: Scalable, private smart contracts,'' in {\em 27th USENIX Security
  Symposium}, (Baltimore, MD), pp.~1353--1370, Aug. 2018.

\bibitem{5158447}
A.~Duminuco and E.~Biersack, ``A practical study of regenerating codes for
  peer-to-peer backup systems,'' in {\em 29th IEEE International Conference on
  Distributed Computing Systems}, (Montreal, Canada), pp.~376--384, 2009.

\bibitem{8845185}
Y.~Savchenko and Y.~Liu, ``Optimizing degree distributions of lt-based codes
  with deep reinforcement learning,'' in {\em IEEE Infocom (Workshop)}, (Paris,
  France), pp.~228--233, May 2019.

\bibitem{XORE13}
M.~Sathiamoorthy, M.~Asteris, D.~Papailiopoulos, A.~G. Dimakis, R.~Vadali,
  S.~Chen, and D.~Borthakur, ``{XORing} elephants: novel erasure codes for big
  data,'' {\em Proc. of VLDB Endowment}, vol.~6, pp.~325--336, Mar. 2023.

\end{thebibliography}
\end{document}